\documentclass[useAMS,usenatbib]{mn2e}
\usepackage{graphicx}
\usepackage{amssymb}
\usepackage{txfonts}
\newcommand{\ee}{$e^\pm$}

\newcommand{\lh}{\ell_{\rm h}}
\newcommand{\ls}{\ell_{\rm s}}
\newcommand{\lth}{\ell_{\rm th}}
\newcommand{\lnth}{\ell_{\rm nth}}
\newcommand{\sax}{{\it Beppo\-SAX}}

\newcommand{\msun}{{\rm M}_{\sun}}
\newcommand{\ledd}{L_{\rm E}}
\newcommand{\xte}{{\textit {RXTE}}}
\newcommand{\pca}{{\it RXTE}/PCA}
\newcommand{\hexte}{{\it RXTE}/HEXTE}

\newcommand{\asm}{{\it RXTE}/ASM}

\newcommand{\integral}{{\it INTEGRAL}}
\newcommand{\chisq}{$\chi^{2}$}
\title[Spectral variability in Cygnus X-3]{Spectral variability in Cygnus X-3} 

\author[L. Hjalmarsdotter et al.]
{L.~Hjalmarsdotter,$^{1,2}$\thanks{E-mail:nea@astro.su.se}
A. A.~Zdziarski,$^{3}$ A.~Szostek,$^{4}$ D.~C.~Hannikainen$^{1,5}$\\
$^{1}$Observatory, PO Box 14, FIN-00014 University of Helsinki, Finland\\
$^{2}$Stockholm Observatory, Department of Astronomy, AlbaNova University Center, 106 91 Stockholm, Sweden\\
$^{3}$Centrum Astronomiczne im.\ M. Kopernika, Bartycka 18, 00-716 Warszawa, Poland\\
$^{4}$Astronomical Observatory, Jagiellonian University, Orla 171, 30-244 Krak\'ow, Poland\\
$^{5}$Mets\"ahovi Radio Observatory, Helsinki University of Technology TKK, Mets\"ahovintie 114, FI-02540 Kylm\"al\"a, Finland}

\pagerange{\pageref{firstpage}--\pageref{lastpage}} 
\pubyear{2008}
\begin{document}
\maketitle
\label{firstpage}

\begin{abstract} 
We model the broad-band X-ray spectrum of Cyg X-3 in all states displayed by this source as observed by the {\it Rossi X-ray Timing Explorer}. From our models, we derive for the first time unabsorbed spectral shapes and luminosities for the full range of spectral states. We interpret the unabsorbed spectra in
 terms of Comptonization by a hybrid electron distribution and strong Compton reflection. We study the spectral evolution and compare with other black hole as well as neutron star sources. We show that a neutron star accretor is not consistent with the spectral evolution as a function of $\ledd$ and especially not with the transition to a hard state. Our results point to the compact object in Cyg X-3 being a massive, $\sim 30\msun$ black hole.
\end{abstract}


\section{Introduction}
Cygnus X-3 is one of the brightest as well as one of the first known X-ray binary systems. It was discovered already 40 years ago (Giacconi et al. 1967) and has been observed by all X-ray missions as well as extensively observed in other wavelengths from radio to PeV energies, although without any confirmed detection above the hard X-ray regime. An interpretation of its intrinsic X-ray properties and spectral evolution is, however, still missing from the literature and the source remains enigmatic. Despite several recent attempts to apply physical models to explain its intrinsic broadband spectrum  (Vilhu et al. 2003; Hjalmarsdotter et al. 2004; Hjalmarsdotter et al. 2008) including a detailed modelling of the wind parameters (Szostek \& Zdziarski 2008), the underlying spectral variability and the true accretion geometry have yet to be explained.  

In this paper, we make use of recent constraints from Hjalmarsdotter et al. (2008), hereafter Hj08 and Szostek \& Zdziarski (2008), hereafter SZ08, to interpret the intrinsic spectral variability observed in Cyg X-3 and discuss its spectral evolution in comparison with other X-ray binaries.  For the first time, we present unabsorbed spectral shapes and luminosities characteristic for the full spectral variability of this source. We use archival data from the {\it Rossi X-ray Timing Explorer\/} ({\xte}) which include the full range of observed spectral states.

\begin{figure}
\includegraphics[width=0.5\textwidth]{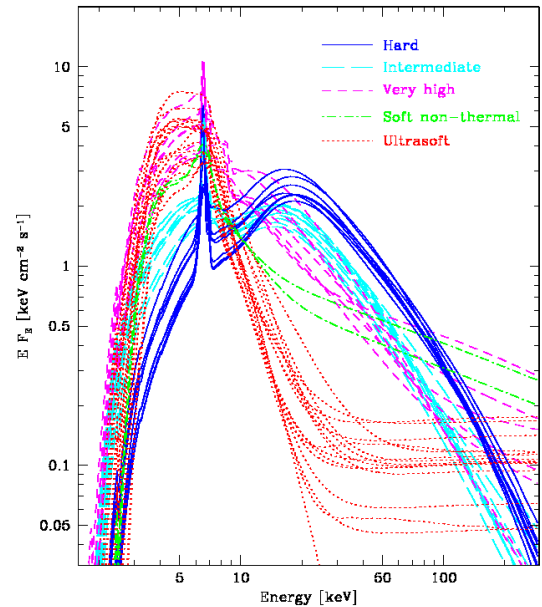} 
\caption{The absorbed spectral shapes of Cyg X-3 from Szostek \& Zdziarski (2004). The 42 observations are averaged into 5 groups, the hard state (blue solid line), the intermediate state (cyan long dashes), the very high state (magenta short dashes), the soft non-thermal state (green dot-dashes) and the ultrasoft state (red dotted line).}
\label{abs}
\end{figure}

%The binary system is very compact, with an orbital period of only 4.8 hours, and enshrouded in the dense stellar wind of the Wolf-Rayet donor, causing strong absorption of the X-ray emission.

%Cyg~X-3 is located at a distance of 9 kpc (Dickey 1983, assuming 8 kpc for the distance to the Galactic centre; Predehl et al. 2000), close to the galactic plane.

\section{The \xte\/ data}
We use the data from 42 pointed observations with the \xte\/ Proportional Counter Array (PCA) and the High Energy X-ray Transient Experiment (HEXTE) between 1996--2000 for which PCA data from Proportional Counter Units 0--2 were available. The PCA spectra were extracted in the 3--25 keV range using only the top layer of the detectors. The corresponding HEXTE spectra were extracted in the 15--110 keV energy band, using both HEXTE clusters. We allow a free relative normalization of the HEXTE spectra with respect to that of the PCA. The observations are listed in Table 1. The data were previously published in Szostek \& Zdziarski (2004). They found that the 42 observations could be divided into five groups based on the shape of the observed absorbed spectrum. Unabsorbed spectral shapes and luminosities were however not discussed in that preliminary study and it was not clear whether the observed differences were due to absorption effects or to intrinsic spectral evolution. The data is also the same as used for the classification of X-ray and radio states in Szostek, Zdziarski \& McCollough (2008).  All the observations and their division into the five groups are shown here in Fig. \ref{abs}. The figure replaces fig.~1 in Szostek \& Zdziarski (2004) and in Szostek et al. (2008) where there is a mismatch between colours and linestyles.

Here, we re-model the average spectra of each group for a detailed investigation of the physics of the observed spectral states, as well as the 42 individual spectra to construct colour-colour and colour-luminosity diagrams. A 1 per cent systematic error was added to each of the averaged spectra. We label the spectra of group 1--5 in Szostek \& Zdziarski (2004) and Szostek et al. (2008), in order of increasing softness, as the hard state, the intermediate state, the very high state, the soft non-thermal state and the ultrasoft state, respectively.

\begin{table}
\centering \caption{The log of the \xte\/ observations. The spectral classes 1, 2, 3, 4 and 5 correspond to the observations (22--23, 25--30), (1--8, 24), (9--13, 19, 21), (41--42), and (14--18, 20, 31--40), respectively. }
\begin{tabular}{rcccl}
\hline
No. & Obs.\ ID & Start (MJD) & End (MJD) & Exposure (s)\\
\hline
  1 &   10126-01-01-00 & 50319.456 & 50319.691 & 8368\\ 
  2 &   10126-01-01-01 & 50321.456 & 50321.632 & 6224\\ 
  3 &  10126-01-01-010 & 50321.151 & 50321.432 & 12832 \\ 
  4 &   10126-01-01-02 & 50322.524 & 50322.763 & 7520\\ 
  5 &  10126-01-01-020 & 50322.258 & 50322.499 & 10512 \\ 
  6 &   10126-01-01-04 & 50323.664 & 50323.832 & 4976\\ 
  7 &   10126-01-01-03 & 50324.665 & 50324.766 & 3328\\ 
  8 &   10126-01-01-05 & 50325.666 & 50325.822 & 4032\\ 
  9 &   20099-01-01-00 & 50495.017 & 50495.308 & 6992\\ 
 10 &   20099-01-01-01 & 50500.285 & 50500.510 & 10960 \\ 
 11 &  20099-01-01-010 & 50500.019 & 50500.259 & 8848\\ 
 12 &  20099-01-01-020 & 50500.765 & 50501.060 & 7040\\ 
 13 &   20099-01-01-02 & 50501.085 & 50501.126 & 1616\\ 
 14 &   20101-01-01-00 & 50604.751 & 50604.846 & 1024\\ 
 15 &   20101-01-02-00 & 50606.006 & 50606.101 & 5696\\ 
 16 &   20101-01-03-00 & 50609.541 & 50609.644 & 1328\\ 
 17 &   20101-01-05-00 & 50612.473 & 50612.576 & 1616\\ 
 18 &   20101-01-06-00 & 50616.762 & 50616.852 & 4800\\ 
 19 &   20101-01-04-00 & 50618.478 & 50618.579 & 864 \\ 
 20 &   20101-01-07-00 & 50624.724 & 50624.835 & 3824\\ 
 21 &   20101-01-08-00 & 50632.690 & 50632.800 & 424\\ 
 22 &   20101-01-09-00 & 50652.630 & 50652.724 & 3760\\ 
 23 &   20101-01-10-00 & 50661.697 & 50661.792 & 4864\\ 
 24 &   20099-02-01-00 & 50717.322 & 50717.498 & 9232\\ 
 25 &   30082-04-02-00 & 50950.950 & 50951.051 & 4080\\ 
 26 &   30082-04-03-00 & 50951.692 & 50951.784 & 3984\\ 
 27 &   30082-04-04-00 & 50952.691 & 50952.783 & 3936\\ 
 28 &   30082-04-05-00 & 50953.692 & 50953.783 & 3600\\ 
 29 &   30082-04-06-00 & 50954.881 & 50954.983 & 4240\\ 
 30 &   30082-04-01-00 & 50949.629 & 50949.717 & 4400\\ 
 31 &   40061-01-02-00 & 51586.378 & 51586.402 & 352 \\ 
 32 &   40061-01-06-01 & 51590.299 & 51590.326 & 1360\\ 
 33 &   40061-01-07-00 & 51590.954 & 51591.060 & 2240\\ 
 34 &   50062-02-03-00 & 51641.043 & 51641.135 & 1920\\ 
 35 &   50062-02-03-01 & 51641.253 & 51641.276 & 1392\\ 
 36 &   50062-02-04-00 & 51642.968 & 51642.996 & 288 \\ 
 37 &   50062-02-06-00 & 51646.028 & 51646.050 & 1264\\ 
 38 &   50062-02-06-01 & 51646.096 & 51646.125 & 880 \\ 
 39 &   50062-02-07-00 & 51648.814 & 51648.851 & 3136\\ 
 40 &   50062-02-08-00 & 51650.038 & 51650.108 & 976 \\ 
 41 &   50062-01-01-00 & 51656.726 & 51656.830 & 2432\\ 
 42 &  50062-01-02-00G & 51661.113 & 51661.214 & 528 \\ 
\hline
\end{tabular}
\label{tab:log}
\end{table}

\section{Spectral model}
\subsection{Continuum model}
We use a Comptonization model, {\tt eqpair} (Coppi 1992, 1999), in which soft seed photons are up-scattered in a hot flow located inside the inner radius of or on top of an accretion disc. The source of the soft seed photons is assumed to be an accretion disc with the spectrum of a disc-blackbody ({\tt diskbb}, Mitsuda et al. 1984, included in {\tt eqpair}). The slope of the Comptonized spectrum is determined by the ratio of heating of the Comptonizing electrons (apart from Coloumb interactions and Compton heating) to cooling due to the injection of soft seed photons. This can be expressed as the ratio of the hard and soft compactnesses $\lh/\ls$ where $\ell \equiv L\sigma_{\rm T}/(r m_{\rm e} c^3)$, $L$ is the luminosity, $\sigma_{\rm T}$ is the Thomson cross section, $r$ is the characteristic size of the X-ray emitting region and $m_{\rm e}$ is the electron mass. The model is weakly dependent on the value of $\ls$, through Coulomb interaction and pair production only, and thus $\ls$ is poorly constrained in the fits. We therefore freeze it at a value of $\ls$=100, corresponding to a high luminosity and a small radius of the Comptonizing plasma, appropriate for the case of Cyg X-3, and similar to the case of GRS 1915+105 (Hannikainen et al. 2005; Zdziarski et al. 2005). 
The plasma optical depth, $\tau_{\rm tot}$, includes a contribution from electrons formed by ionization of the atoms in the plasma, $\tau_{\rm e}$, as well as a contribution from \ee pairs, $\tau_{\rm tot}-\tau_{\rm e}$. The electron distribution, including the temperature, $T_{\rm e}$, is calculated self-consistently from energy balance and can be purely thermal or hybrid if an acceleration process is present. The acceleration rate is assumed to have a power-law shape with index $\Gamma$ for Lorentz factors between 1.3 and 1000. In the case of a hybrid plasma $\lh=\lth+\lnth$, i.e., the thermal heating plus the power supplied to accelerated electrons. This model for the continuum intrinsic spectrum is the same as used by Vilhu et al. 2003, Hjalmarsdotter et al. 2004, Hj08 and SZ08, and its application to Cyg X-3 is discussed more elaborately especially in Hj08. The model is also described in detail by Gierli\'nski et al. (1999). 

Compton reflection is included ({\tt pexriv}, Magdziarz \& Zdziarski 1995), parametrized by a solid angle $R=\Omega/2\pi$. In an un-obscured geometry $R$ corresponds to the true solid angle of the reflector as seen from the primary X-ray source. If the primary source is obscured, $R$ also involves the relative amount of emission seen in reflection to that observed directly, and can thus greatly exceed 2, the maximum value for an isotropic source in an un-obscured situation (see further discussion in Hj08 and SZ08). The strength of the reflected component also depends on the system inclination, which is not well known in Cyg X-3. The strong orbital modulation suggests a high inclination, but not too high since there is no eclipse by the secondary. For the purpose of modelling the reflection, we use $60\degr$ (Vilhu et al. 2007). We allow the reflecting medium to be ionized but find that our fits are consistent with a neutral reflecting medium.

\subsection{Absorption effects in the wind}
Absorption in Cyg~X-3 is a complex matter. Studies with e.g. {\it Chandra} (Paerels et al. 2000) have revealed a rich line complex with both absorption and emission features at energies $<$10 keV. SZ08 recently studied the effect of the stellar wind from the Wolf-Rayet companion on the spectra of Cyg X-3 using medium resolution \sax\/ data and modelled the wind using the photoionization code {\tt cloudy} (Ferland et al. 1998). They found that the emission and absorption structure at low energies required a two-phase wind made up by a hot tenuous plasma together with cool dense clumps filling $\sim 1$ per cent of the wind volume. Since the \xte\/ data lack high spectral resolution and cover only energies above 3 keV, and since we are more interested here in the intrinsic spectra than the wind parameters, we use a simplified model for the absorption based on the findings of SZ08. 

In a first attempt aiming to reproduce the physical conditions of the wind as well as possible, although in a simplified approach, we model the wind using {\tt absori} (Done et al. 1992), with a weakly ionized one-temperature plasma with initial values for the optical depth, temperature and ionization parameter set to average values from the results in SZ08. We find that freezing the values for the temperature and ionization parameter to the average values from SZ08 does not give satisfactory fits to any of the data sets, showing that a one-phase single-temperature and single-ionization absorber is not a good approximation to the complex wind structure. However, allowing column density, temperature and ionization free in our fits, we find that our data can not constrain these values or break the degeneracy between them. In most cases, the obtained error ranges are consistent with neutral absorption.

On the other hand, we find that, just like in Vilhu et al. 2003, Hjalmarsdotter et al. 2004 and Hj08, very good fits to the data can be obtained using a two-fold neutral absorber ({\tt absnd}) with one medium with column density $N_{\rm He, tot}$, covering the entire source and one medium with column density $N_{\rm He, par}$, covering a fraction $f_{\rm c}$, of the source (with the column densities expressed in He rather than H, given that Wolf-Rayet stars contain very little hydrogen, see further discussion on abundances below). This model was also used by SZ08 as a first approximation for the effect of the wind absorption in order to calculate the ionizing spectrum for the wind modelling. The overall similarity of the obtained intrinsic spectral shapes in that first step to those obtained in the final wind modelling in SZ08, shows that this two-fold model, although more phenomenological than physical, gives a very good approximation to the effect of the wind absorption in Cyg X-3. In this paper, where we aim at finding physically consistent models for the intrinsic spectral shapes rather than any detailed modelling of the absorbing medium, we therefore use the neutral two-fold absorber to model the continuum absorption since it gives a better phenomenological approximation of the true wind effects than any simplified ionized absorber. We find that for our purposes, this is a sufficiently accurate approach, but note that the values of the column density can not really be constrained based on our data.  

We first fit the data with the abundances of a Wolf-Rayet star of WNE type from Hamann \& Gr\"afener (2004), the same abundances as used by SZ08. Since our simplified modelling of the absorber does not take into account ionization, it is, however, not consistent with high iron abundances for column densities close to the ones fitted by SZ08.
Furthermore, we find that for some spectral states where reflection plays an important role, especially in the hard state (see Section 4.1) the \xte\/ data do not seem to be consistent with high iron and metal abundances. We thus use normal solar abundances (Anders \& Ebihara 1982), but with the metal, $A_{>\rm He}$ (only in {\tt eqpair}), and iron, $A_{\rm Fe}$, abundances (requiring the same value for the absorption as in {\tt eqpair}) as free parameters. The abundances are thus allowed to differ between different spectral states. This does not correspond to a physical reality, but is rather a compensation of not treating ionization, which may well differ between the different spectral states. We note that an anomalously low iron abundance, much lower than the values in our best fit models, were in fact suggested by Terasawa \& Nakamura (1994, 1995). We give column densities of the local absorber as re-scaled to helium as $N_{\rm He}$. 

In Hj08, an additional Thomson absorber was used to account for additional attenuation caused by electron scattering in the stellar wind, presumably responsible for the strong orbital modulation. The phase averaged data was corrected by increasing the normalization by 1/0.75 times to roughly correspond to phase maximum. Here, we reconsider this assumption. If the wind is highly ionized  scattering does not remove photons so the photons missing at phase minimum have to show up somewhere else. The simplest hypothesis is that they get back-scattered, and show up around phase maximum. This means that when we look at the source at the phase where the optical depth is lowest, we see photons back-scattered from the direction where the optical depth is the highest. Averaging over the phase, we see {\it all} the photons, and no correction is required. We do not know the exact degree of ionization of the scattering medium. The wind-modelling in SZ08 suggests, however, that the hot tenous highly ionized phase of the wind makes up 99 $\%$ of its volume. We thus assume that photons are not significantly removed in the scattering process and do not make any corrections to the derived luminosity due to scattering. This gives the most conservative estimate for the luminosity and we note that the actual luminosity may be higher.     

Finally, we fix the interstellar absorber to $N_{\rm H, gal}=1.5\times10^{22}$ cm$^{-2}$ (e.g. Chu \& Bieging 1973).

\subsection{The iron complex}
A strong edge at 7 keV apparent in both data sets modelled in SZ08 implies the presence of neutral iron, not accounted for in their model of the wind. This confirms the presence of at least some additional cool material in the system, either in the form of a reflector, or as an absorber in addition to the Wolf-Rayet wind, or both. In our models to the \xte\/ data, the edge results from a combination of neutral absorption and reflection.   
An additional edge at $\sim$9 keV is also required by the data. In the wind modelling by SZ08, this edge arises from highly ionized iron in the otherwise optically thin hot inter-clump wind medium. For simplicity, we model this hot phase with an edge with energy $E_{\rm edge}$ and optical depth $\tau_{\rm edge}$. We find our values of the optical depth of this edge to correspond well to the column density of the hot phase in SZ08.

The presence of a range of ionization species of iron in the Cyg X-3 wind also gives rise to a rich emission line spectrum, resolved by {\it Chandra} (Paerels et al. 2000) and partly resolved by \sax\/ (SZ08). The iron line complex is also clearly visible in the \xte\/ data, but due to the low spectral resolution of the PCA instrument, a single broad gaussian centered at $\sim6.5$ keV is sufficient to model it. For a more physical approach, however, we separate the most prominent wind emission line at 6.7 keV, as observed by {\it Chandra} and \sax\/, from any possible additional flourescent line arising from either absorption or reflection. In the case of strong reflection with $R\ge 1$ an accompanying flourescent line is indeed expected. We thus model the iron line complex with two gaussians, one with a fixed energy  $E= 6.7$ keV and fixed physical width $\sigma=0.1$ (corresponding to an arbitrary thin line with the PCA resolution) and allow an additional line with energy $E_{\rm Fe}$ between 6.4 and 7.5 keV, physical width $\sigma_{\rm Fe} \leq 0.4$ keV and equivalence width $EW_{\rm Fe}$ We find that including even more line structure, e.g. a line at 6.9 keV, also seen in higher resolution data, is not required in our fits to the \xte\/ data.  

We note that the large error ranges on most of our best-fit parameters in Table 2 indicate that they are not at all very well constrained. Some parameters, e.g. the column density $N_{\rm He}$, can simply not be constrained by our data. For several of the other parameters, however, the large error range is due to the presence of several, themselves much more narrow, local minima. Including all these local minima results in a very large global error range. We have found that freezing the blackbody temperature at the best-fit temperature when determining the error ranges for the other parameters reduces the number of local minima, especially those with intrinsic spectral shapes and luminosities far from the best fit model, and especially in the hard and intermediate states. This procedure was followed for all states except the ultrasoft state. The problem with very large error ranges and unconstrained column densities, however, remains. We also can not fully rule out the presence of additional minima in this very complicated multidimensional parameter space.

\section{Spectral states}
\subsection{The hard state}
\begin{figure*}
\includegraphics[width=0.47\textwidth]{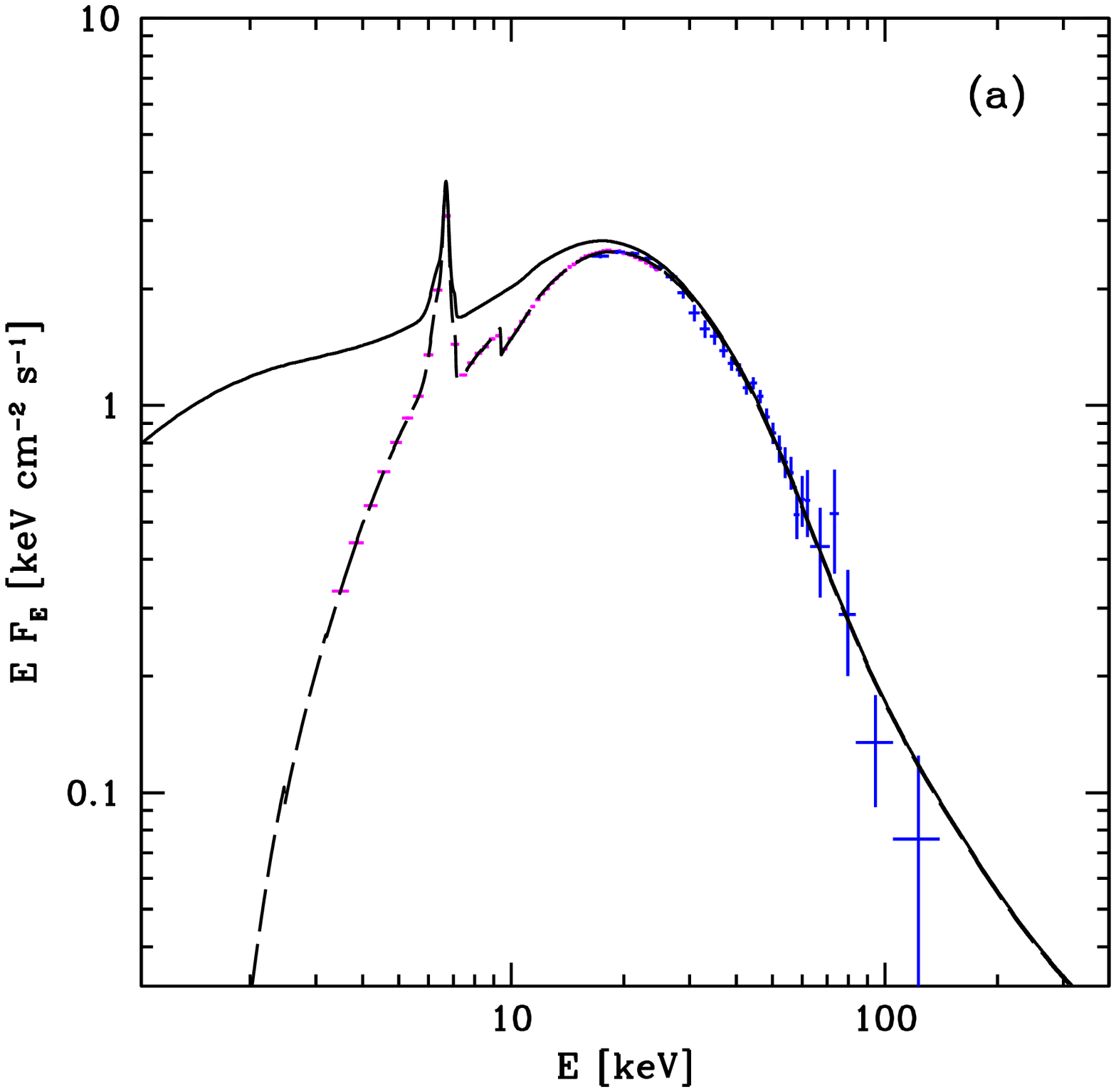} 
\includegraphics[width=0.47\textwidth]{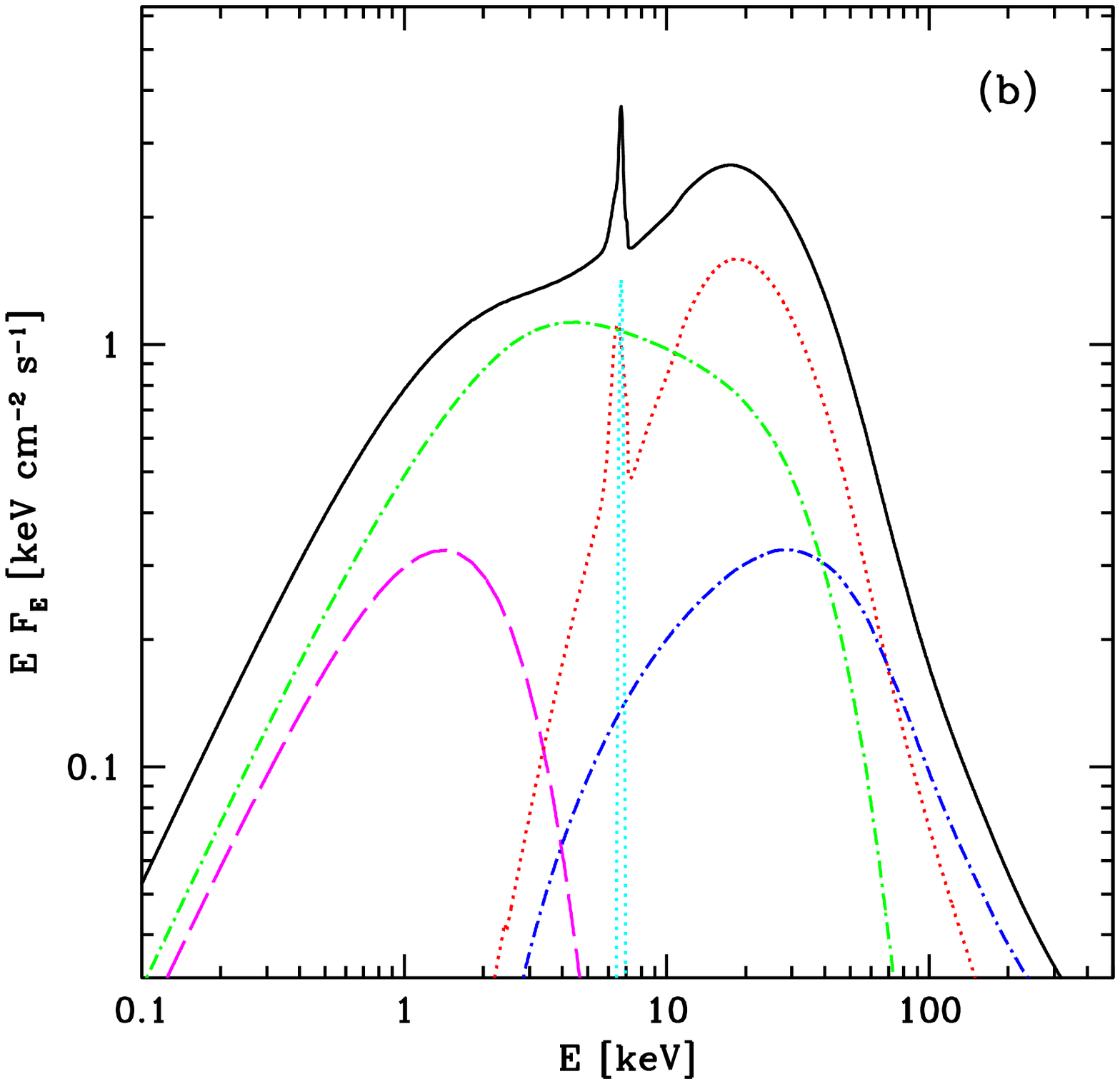} 
\caption{{\it a:} Averaged data from eight observations with the \pca\/ (magenta) and \hexte\/ (blue) of Cyg~X-3 in its hardest state together with the model described in Section 4.1, absorbed (dashed) and corrected for absorption (solid line). {\it b:} The components of the unabsorbed model: the unscattered blackbody (magenta long dashes), Compton scattering from thermal (green short dashes) and non-thermal (blue dot-dashes) electrons, reflection and the Fe K line (red dotted line) and the Fe 6.7 keV emission line from the wind (cyan dotted line).}
\label{hard}
\end{figure*}

The hardest state of Cyg~X-3 differs from the canonical hard state, as seen in e.g. Cyg~X-1, in that it peaks at much lower energies, $\sim20$ keV, compared to other black holes as well as neutron stars in their hard states. Such a low cut-off implies an electron temperature of the Comptonizing electrons below 10 keV, usually observed only in soft states where strong cooling is provided by the massive influx of soft seed photons from a strong blackbody component. In addition, it was found by Hj08 that any attempt to fit the data with thermal Comptonization proved difficult, since the high-energy slope does not match well a cut-off due to the finite temperature of a Comptonizing electron distribution (similar to but not equal to an e-folded power-law). Hj08 showed that the Cyg~X-3 hard state could indeed be fitted with a model with a high amplification factor ($\lh/\ls>7$), despite the low cut-off energy, if the power to the electrons in the plasma was provided entirely by acceleration. A model where the apparent hard state was instead only a result of increased absorption in the stellar wind also gave a good fit to the \integral\/ data in Hj08. The best fit in that study was found for a model dominated by Compton reflection, where most of the direct emission was obscured from view. 

The \xte\/ data of Cyg X-3 in its hard state used here consist of the spectra from 8 observations, made between July 1997 and August 1998. The variations between them are small and the individual spectra as well as the averaged one is similar to the hard state as observed by \integral\/ in Hjalmarsdotter et al. (2004) and in Hj08. For the sake of statistics, we derive our best fit models based on the averaged spectrum of all 8 observations. 

We find that all three models tested in Hj08 give good fits ($\chi^2/ \nu <1.2$) also to the \xte\/ data, with the reflection dominated model again providing the best fit ($\chi^2/\nu=0.48$) (We note that our very low \chisq\/ values for the \xte\/ data may well be due to an overestimation of the systematic errors added to our data). Despite its ability to fit the data, this model was however ruled out as a physical description of the hard state of Cyg X-3 by Hj08, since it, just like the wind absorption model, presented in the same paper, required a very luminous soft incident spectrum. This is not in agreement with the other results of that study which strongly preferred a `real' state transition rather than obscuration effects (see further discussion in Section 6.1). In SZ08, \sax\/ data from a hard state was modelled with a continuum model with equally strong reflection ($R=10$) as in the reflection model of Hj08 but with a harder incident spectrum with a very high blackbody temperature and almost completely non-thermal distribution of the Comptonizing electrons. Such a spectrum does not give a satisfactory fit to the \xte\/ data with $\chi^2/\nu \sim2$. %(We note that the hard state spectrum in SZ08 is not really a «hard' spectrum in terms of the balance of heating versus cooling as $\lh/\ls=0.7$, and the hard slope is merely a result of the high blackbody temperature assumed where the hard slope is just the $\nu\propto?$ part of the disc blackbody spectrum).

However, we find a very good fit ($\chi^2/\nu=0.44$) for a model with moderately strong reflection and with an incident spectrum of moderate hardness and luminosity.
The model is plotted together with the data in Fig. \ref{hard}a, and the components of the unabsorbed model are shown in Fig. \ref{hard}b. The best fit parameters are listed in Table 2. 

\begin{figure*}
\includegraphics[width=0.47\textwidth]{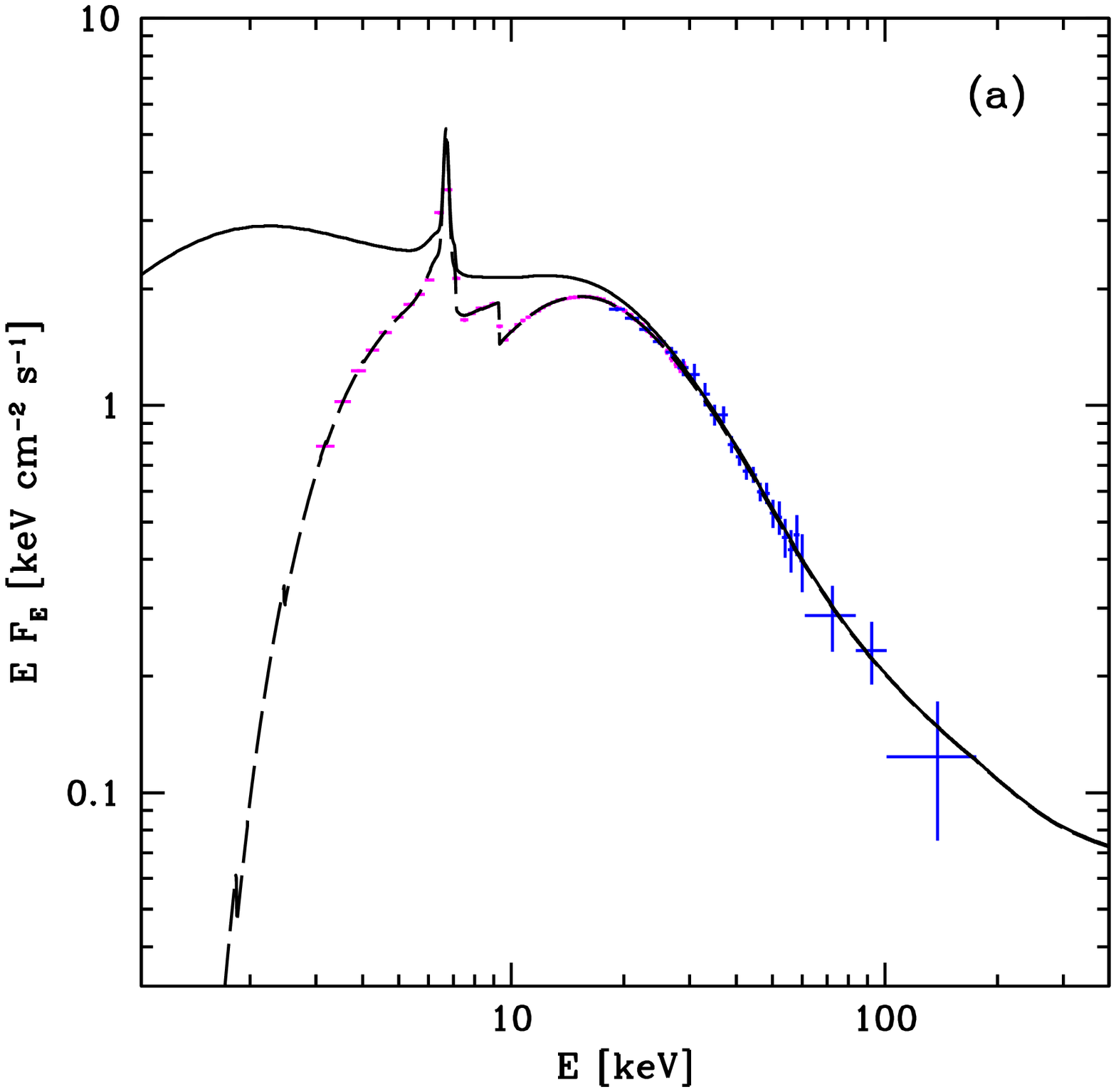} 
\includegraphics[width=0.47\textwidth]{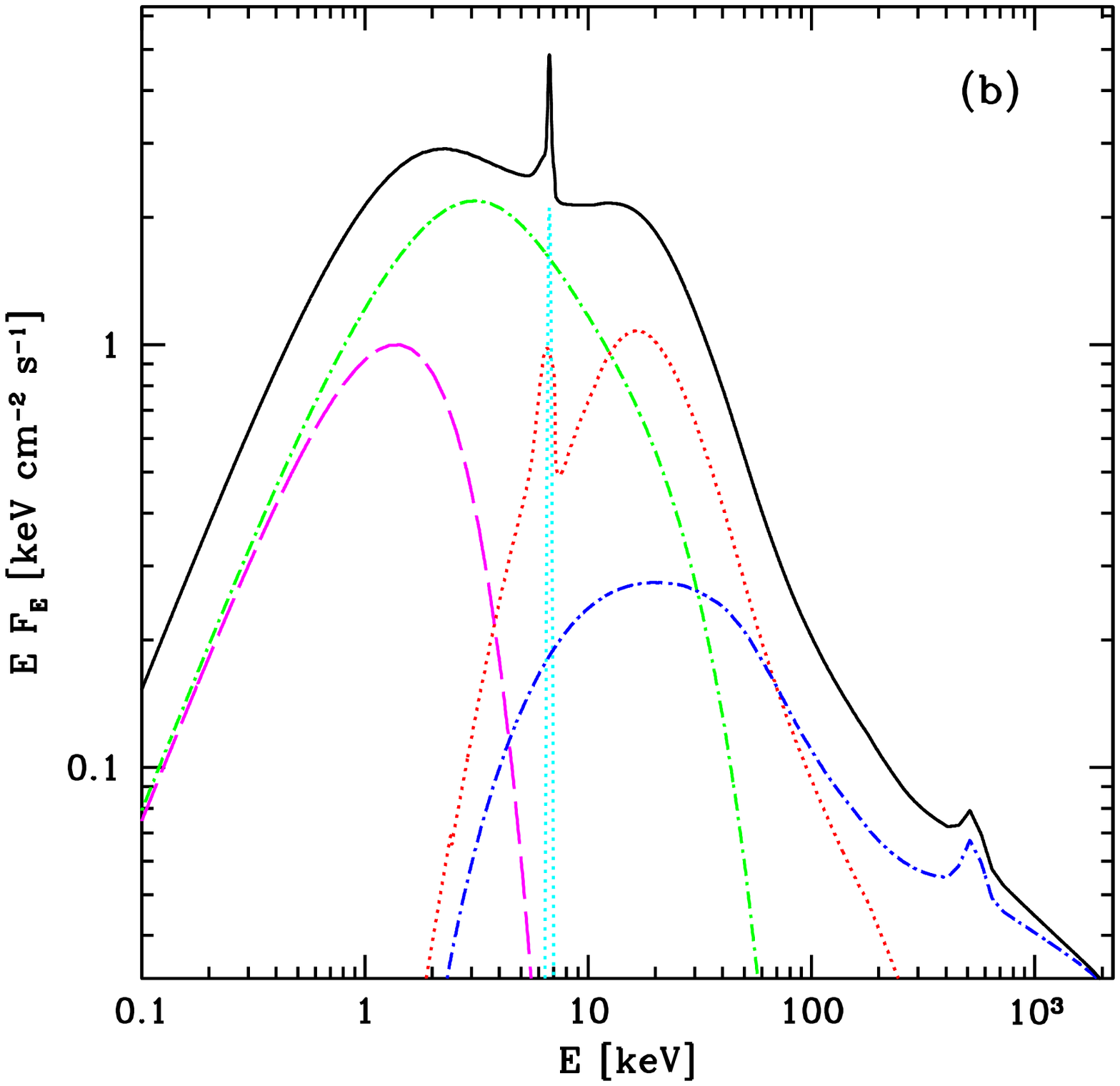} 
\caption{{\it a:} Averaged data from 9 observations by \pca\/ (magenta) and \hexte\/ (blue) of Cyg~X-3 in its intermediate state, together with the model described in Section 4.2, absorbed (dashes) and corrected for absorption (solid line). {\it b:} The components of the unabsorbed model: the unscattered blackbody (magenta long dashes), Compton scattering from thermal (green short dashes) and non-thermal (blue dot-dashes) electrons, reflection and the Fe K line (red dotted line) and the Fe emission line at 6.7 keV (cyan dotted line) from the wind.}
\label{int}
\end{figure*}

In our model, the observed spectrum above 10 keV is dominated by strong Compton reflection that creates the characteristic shape of the cut-off around 20 keV. The model requires $R=2.5$, which corresponds to a situation just exceeding maximum possible reflection in a geometry where most of the direct emission is still observed directly (for $i=60\degr$). 

The incident spectrum is that of a moderately hard state. With $\lh/\ls$=2.0, the power supplied directly to the electrons is higher than the power in the irradiating seed photons, but due to the low electron temperature of only 12 keV, the Comptonized spectrum (the sum of the thermal and non-thermal Comptonized components) before reflection is flat in $\nu F\nu$ space. The best-fit blackbody temperature of 0.6 keV suggests that the disc is truncated at some radius far from the innermost stable orbit, but not as far away as in the canonical hard state of e.g. Cyg~X-1 (Gierli\'nski et al. 1997), or as suggested by the non-thermal model in Hj08. We note, however, that the exact blackbody temperature, in our proposed model located at energies much lower than the \xte\/ bandpass at 3 keV, is not well constrained by the data.

In addition to the Fe {\sc xxv} line at 6.7 arising in the wind, an additional broader line centered at 6.4 keV is required by the data, in agreement with strong reflection from a neutral medium. We note that the fitted $EW_{\rm Fe}$ of the broad line is almost twice as high as expected (calculated self-consistently using {\tt thcompfe} Zdziarski, Johnson \& Magdziarz 1996; Zycki, Done \& Smith 1999) from the fitted $R$ using the best fit abundances. Using the WR-abundances from Hamann \& Gr\"afener (2004), however, the fitted line $EW_{\rm Fe}$ agrees well with the expected $EW_{\rm Fe}$ for both the hard and the intermediate state.

It is clear that even if the accretion geometry is similar to that of the canonical hard state, the physics of this accretion flow differs from that of a standard optically thin advective flow where $\tau\sim$1 and the electron temperature is observed to be between 50--100 keV for black holes (e.g. Zdziarski \& Gierli\'nski 2004), and 40--60 keV for neutron stars (Barret 2001). The optical depth of $\tau=4.8$ includes a small fraction of pairs. A 38 per cent non-thermal fraction of the electron distribution with injection with power-law index 3.9 is required to fit the highest energy HEXTE data. We note that the HEXTE spectrum above 100 keV does not agree with that observed in the hard state by SPI in Hj08 (fig.\ 8 in that paper) but agrees with the ISGRI data (same figure). According to the model presented here, Cyg~X-3 is not likely to be detected by {\it GLAST} in its hard state, unless the maximum Lorentz factor of the accelerated electrons is larger than the assumed 1000.

We note that a slightly softer version of our moderate reflection model gives a rather good fit to the \sax\/ data of SZ08, and a slightly better fit than the phenomenological model presented there. This, however, only holds if the abundances are allowed to, as here, differ from normal Wolf-Rayet ones, which they were not in SZ08, and which may not, as discussed in Section 3.1, correspond to a physical reality. We have also found this model to give a good fit to the \integral\/ data, better than the wind-absorption and the non-thermal models in Hj08, but not as good as the reflection model presented there. The \integral\/ data are indeed much more difficult to model in part due to the large discrepancy between the SPI and ISGRI data in Hj08. Since the model presented here for the hard state is also agreement with the results from Hj08 in that it represents a state with the Compton amplification factor $\lh$/$\ls>1$, and probable truncation of the inner disc, we suggest it as a better physical description of the intrinsic spectrum of Cyg~X-3 in its hard state than any earlier presented models.

Using our best-fit model of the averaged data as an input, we also model the eight individual spectra to derive their intrinsic shapes and colours (see Section 5). Due to uncertainty of whether the small variations between the individual spectra is due to intrinsic spectral variability or to a change in the absorbing medium, we, however, refrain from discussing and/or interpreting this intra-state variability in terms of change in the parameter values. The same procedure is performed for all five spectral states.

\subsection{The intermediate state}
An intermediate state generally refers to a spectral state in between the hard and the soft state and is usually observed in transitions. In Cyg X-3, state transitions are frequent and an intermediate state should therefore be rather common. The \xte\/ data presented here consist of 9 spectra, one group of continuous observations in August 1996, judging from the ASM lightcurve just before a transition from a hard to a soft state, plus one observation from September 1996 after a short transition into a hard state and back to a soft. %Radio levels were quiescent or comparable to those observed in the hard state.

Our best fit model to the Cyg~X-3 intermediate state is similar to that of the hard state, with a similar blackbody temperature, but with heating of the electrons balanced by cooling by seed photons to give $\lh/\ls=0.96$. The main difference in the observed spectral shape is the somewhat lower amplitude of reflection than in the hard state. A harder injection spectrum of the non-thermal electrons also gives more power at energies above a few hundred keV. The bolometric X-ray luminosity is somewhat higher than in the hard state. The model is shown together with the data in Fig. \ref{int}a, and its components in Fig. \ref{int}b. The best fit parameter values are listed in Table 2. 

We note that our model predicts a strong annihilation feature. The actual presence of such a feature (at 511 keV, above the fitted range here) depends on the unknown compactness ($\ls$), and will be weaker at lower compactnesses. In addition, a lower value of the maximum Lorentz factor of the accelerated electrons would produce less high-energy photons and thus less pairs at a given compactness, thus reducing the feature.

\subsection{The very high state}
The term `very high state' is used in the literature to describe a variety of spectral shapes, not without some confusion. It usually describes a spectral state with the presence of both a strong blackbody component and a strong component from thermal Comptonization, typically with the Comptonized component starting already at the peak of the blackbody (e.g. Done \& Gierli\'nski 2004). In some sources the spectral shape in this state is often more or less identical to the intermediate state (as pointed out by Zdziarski \& Gierli\'nski 2004), and differs only in amplitude. The state may or may not be the most luminous one.

\begin{figure*}
\includegraphics[width=0.47\textwidth]{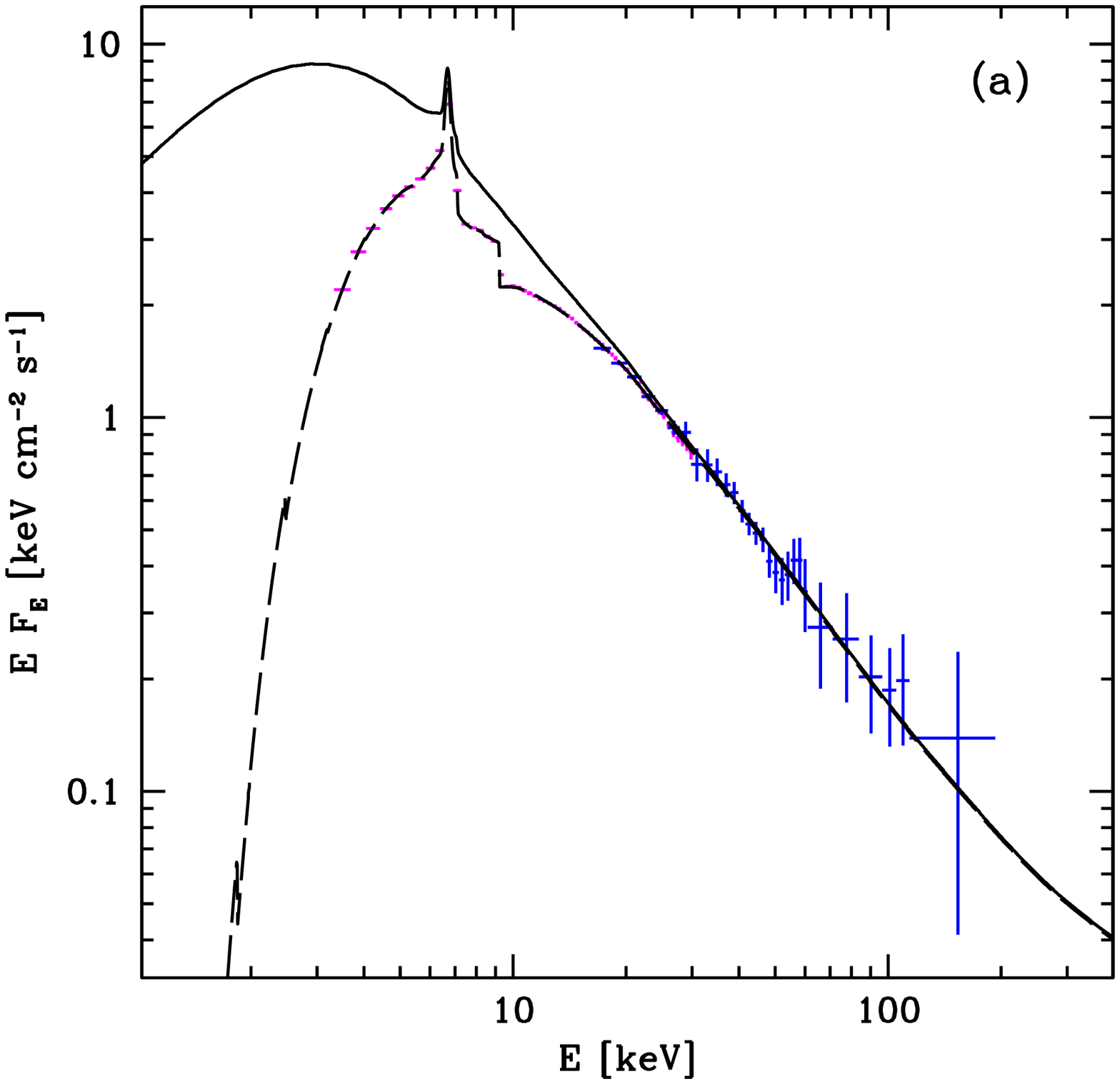} 
\includegraphics[width=0.47\textwidth]{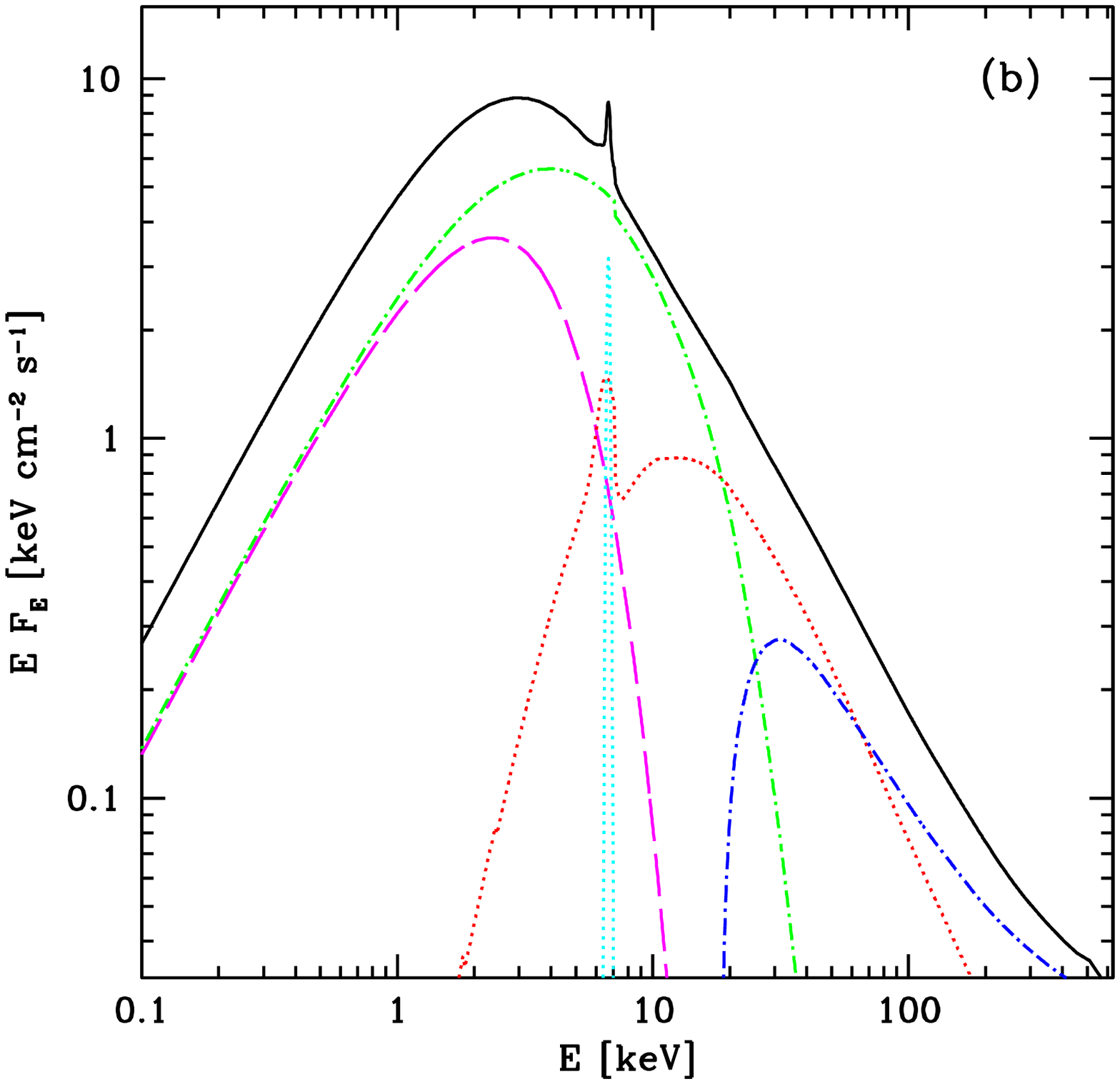} 
\caption{{\it a:} Averaged data from 7 observations  by \pca\/ (magenta) and \hexte\/ (blue) of Cyg~X-3 in its very high state, together with the model described in Section 4.3, absorbed (dashes) and corrected for absorption (solid line). {\it b:} The components of the unabsorbed model: the unscattered blackbody (magenta long dashes), Compton scattering from thermal (green short dashes) and non-thermal (blue dot-dashes) electrons, reflection and the Fe K line (red dotted line), and the Fe 6.7 keV emission line from the wind (cyan dotted line).}
\label{vhigh}
\end{figure*}

The very high state as displayed by Cyg X-3 is a rather soft variant of this state with a strong Comptonized component, rather different from its intermediate state. This very high state is, judging from both the \asm\/ lightcurve (high flux and moderate hardness) and pointed observations by both \xte\/ and \integral\/, the second most common spectral state in Cyg X-3 after the hard state (Hj08). Fig. \ref{vhigh}a shows our best-fit model together with the data averaged from 7 observations by \xte\/ in February, June and July 1997. The spectrum is similar to that in Vilhu et al. (2003) but modelled here in a somewhat different way. In our interpretation, its unabsorbed spectral shape is similar to that of the softer states of GRS~1915+105 and with similar parameters (e.g. Zdziarski et al. 2005). The blackbody temperature is 1.0 keV, and a strong disc component is present in the spectrum. Cooling is strong with $\lh/\ls=0.33$ but the spectrum is still dominated by the Comptonized component. The plasma has a temperature of 5 keV and an optical depth of 3.4. Reflection is strong also in this state, $R=2.3$, but since the spectrum is much softer, reflection does not have as strong an effect on the observed spectral shape. As opposed to the hard and intermediate state, in this very high state the broad iron line has an $EW_{\rm Fe}$ somewhat lower than that expected from fluorescence corresponding to the strength of the continuum reflection component (for both the fitted abundances and for typical WR-abundances). The \xte\/ data of Cyg X-3 in its very high state is consistent with a rather steep distribution of non-thermal electrons and less emission in the model at energies $>100$ keV. The very high state is the most luminous state in Cyg X-3, calculated from the absorbed as well as the absorption corrected bolometric X-ray flux. The inferred geometry is a disc extending close to the innermost stable orbit with the hot Comptonizing flow formed on top of the disc rather than within its inner radius. 

\begin{figure*}
\includegraphics[width=0.47\textwidth]{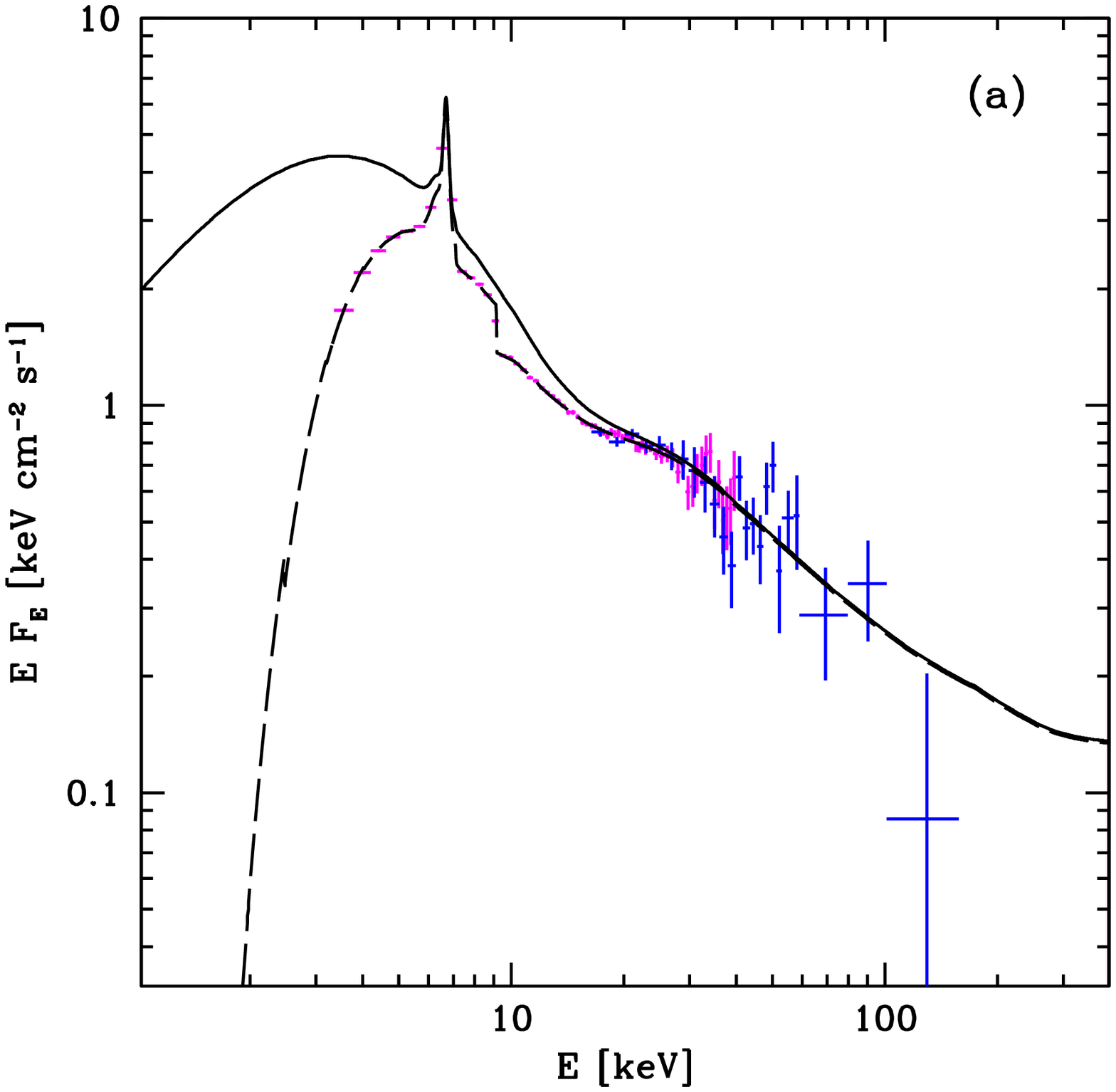} 
\includegraphics[width=0.47\textwidth]{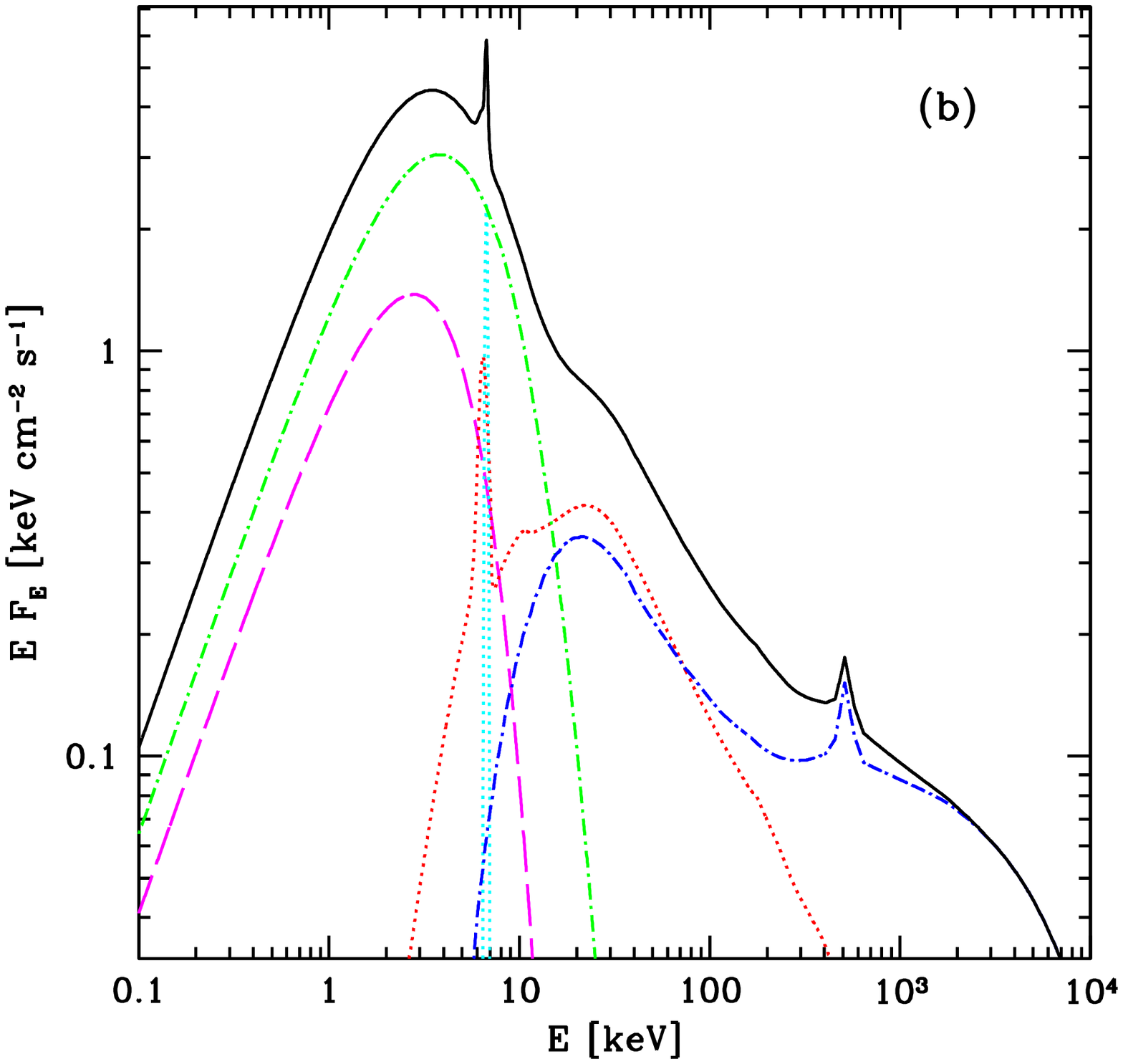} 
\caption{{\it a:} Averaged data from 2 observations by \pca\/ (magenta) and \hexte\/ (blue) of Cyg~X-3 in the state labelled the soft non-thermal state together with the model described in Section 4.4, absorbed (dashes) and corrected for absorption (solid line). {\it b:} The components of the unabsorbed model: the unscattered blackbody (magenta long dashes), Compton scattering from thermal (green short dashes) and non-thermal (blue short dashes) electrons, reflection and the Fe K line (red dotted line) and the Fe 6.7 emission line from the wind (cyan dotted line).}
\label{high}
\end{figure*}

\begin{figure*}
\includegraphics[width=0.47\textwidth]{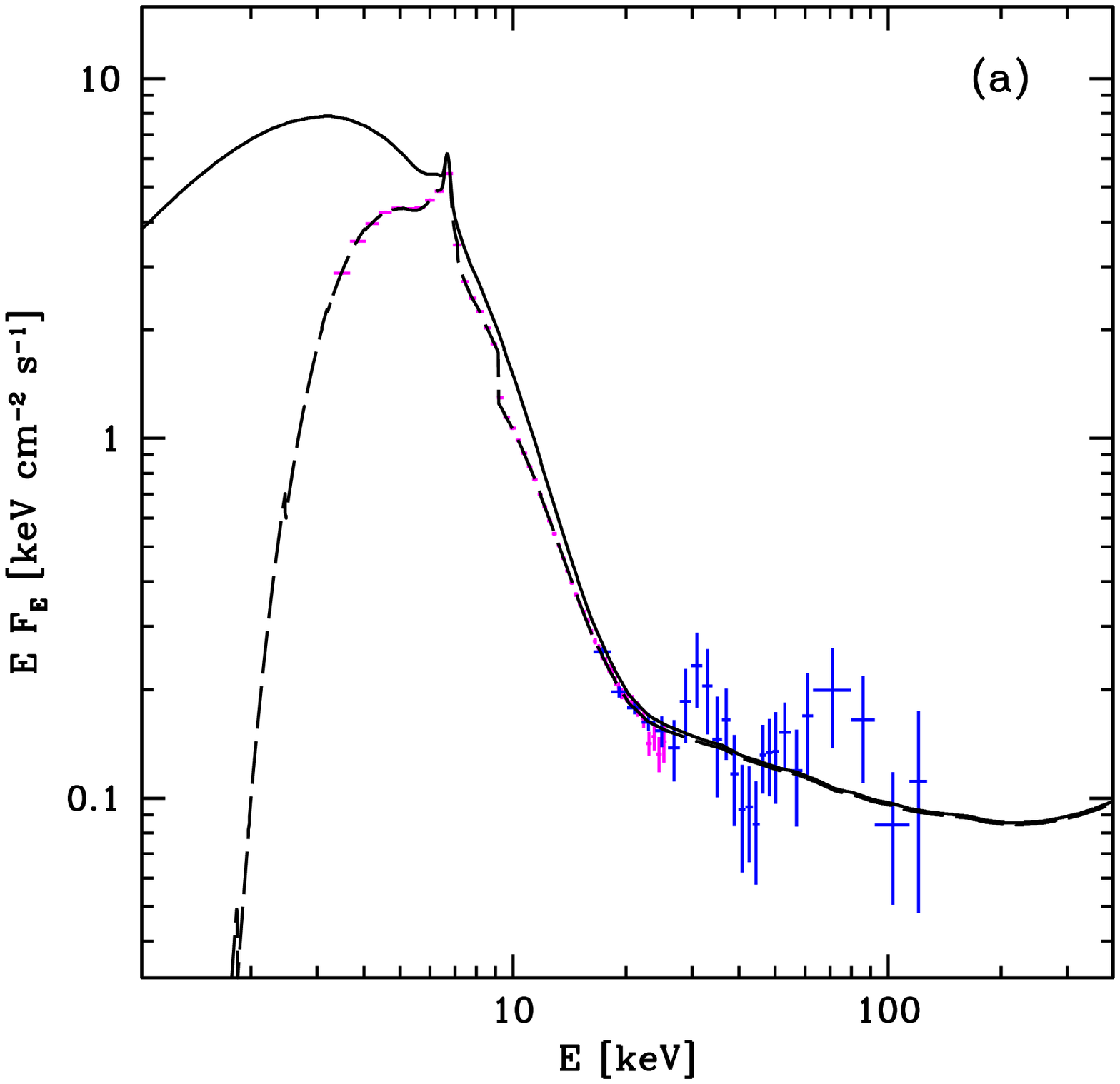} 
\includegraphics[width=0.47\textwidth]{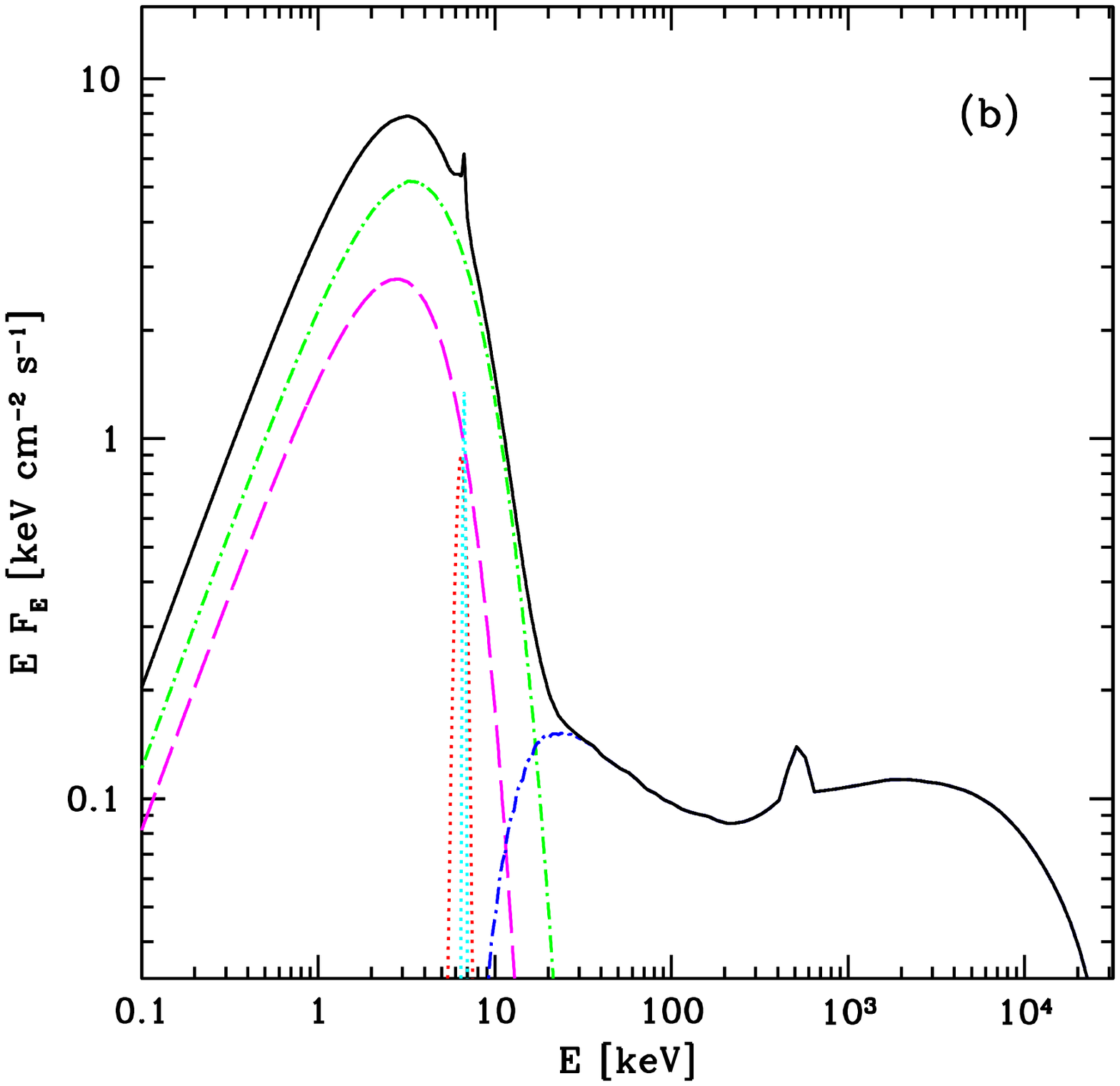} 
\caption{{\it a:} Averaged data from 16 observations  by \pca\/ (magenta) and \hexte\/ (blue) of Cyg~X-3 in its ultrasoft state, together with the model described in Section 4.5, absorbed (dashes) and corrected for absorption (solid line). {\it b:} The components of the unabsorbed model: the unscattered blackbody (magenta long dashes), Compton scattering from thermal (green short dashes) and non-thermal (blue dot-dashes) electrons, the Fe 6.7 line (cyan dotted line) and an additional broad Fe line (red dotted line).}
\label{usoft}
\end{figure*}

\subsection{The soft non-thermal state}
Cyg X-3 also displays a soft state with a moderate disc component and a very strong nonthermal tail.
This state was observed in April 2000, right after a major radio flare. The spectral state is similar to that labelled `high' state in Done \& Gierli\'nski (2004, fig.\ 1a), and appears to be a transitional state between the ultrasoft and the very high states. In terms of luminosity, however, this state in Cyg X-3 is equal to that of the intermediate state and thus the least luminous of the observed soft states in this source.

Our modelling of this state includes a Comptonized blackbody similar to that of the very high state, but weaker, plus a significant non-thermal component requiring an almost 75 per cent non-thermal distribution of the Comptonizing electrons. The model is shown in Fig. \ref{high}. The best fitting parameters are given in Table 2.

\subsection{The ultrasoft state}
The softest state of Cyg~X-3 belongs to the so called ultrasoft state, a state with the soft component dominating the spectrum and with only weak emission above $\sim$20 keV. Such a spectrum is believed to arise at high accretion rates where the thin disc extends all the way into the innermost stable orbit of the black hole (or neutron star). A strong non-thermal tail, with no apparent cut-off below 200 keV, is also present. The ultrasoft state is generally observed at accretion rates ranging from about 0.1$\ledd$ up to $\ledd$ or higher (e.g. Gierli\'nski \& Done 2004) while being absent in sources not reaching luminosities close to 0.1$\ledd$, like e.g. Cyg X-1. The state is in some sources the most luminous one. 

\xte\/ has observed Cyg~X-3 in its ultrasoft state on several occasions, typically right before a major radio flare. Here we model the spectrum averaged from a total of 16 observations in June 1997, February 2000 and June 2000. Fig. \ref{usoft}a, shows the data together with the best fit model with parameters in Table 2. In our unabsorbed version, the spectrum of Cyg~X-3 is indeed very similar to the ultrasoft spectra of e.g. GRS 1915+105, XTE J1550-564 and GX 339-4 (see a comparison of the {\it absorbed} spectral states in fig.\ 8 in Zdziarski \& Gierli\'nski 2004). Fig. \ref{usoft}b shows the components of the model. The spectral de-composition shows that the Cyg~X-3 spectrum in this state is not really blackbody dominated, but dominated by the component arising from thermal Comptonization of the strong blackbody by electrons of moderate temperature. This is similar to the ultrasoft state of GRS~1915+105 in Zdziarski et al. (2005). Just like in GRS~1915+105, simple modelling with a blackbody + power-law model of this state would thus lead to an overestimation of the blackbody temperature. Our best-fit model includes a nearly flat non-thermal tail. If such a tail is real it {\it may} be strong enough to be detected by {\it GLAST}/LAT after one year of operations (according to the {\it GLAST}/LAT performance web page\footnote{{www-glast.slac.stanford.edu/software/IS/glast\_lat\_performance.htm}}). 

Our model is consistent with zero reflection. Models with high reflection, $R\sim5$, are, however, found within $\Delta$\chisq\/ just above 3.0. We note that the \sax\/ data modelled in SZ08 do not seem to be consistent with zero reflection, at least not without a neutral absorbing medium in addition to the ionized wind in their model. The highly reflection dominated model presented in SZ08 is, however, problematic since the strong reflection is not accompanied by any fluorescent Fe line apparent neither in their \sax\/ data, nor seen in e.g. {\it Chandra} data in the Cyg X-3 ultrasoft state (M. McCollough, in prep.) 
In the model presented here, it seems on the other hand as if a broadened line, in addition to the narrow line at 6.7 keV {\it is} required by the data, despite the zero reflection, unless the 6.7 keV line itself is allowed to be broadened. We conclude that due to the low resolution of the PCA instrument, the data used here do not allow for any detailed line diagnostics, but that our zero-reflection model for the continuum in the ultrasoft state agrees with the {\it Chandra} data. 

\begin{figure*}
\includegraphics[width=0.5\textwidth,height=9cm]{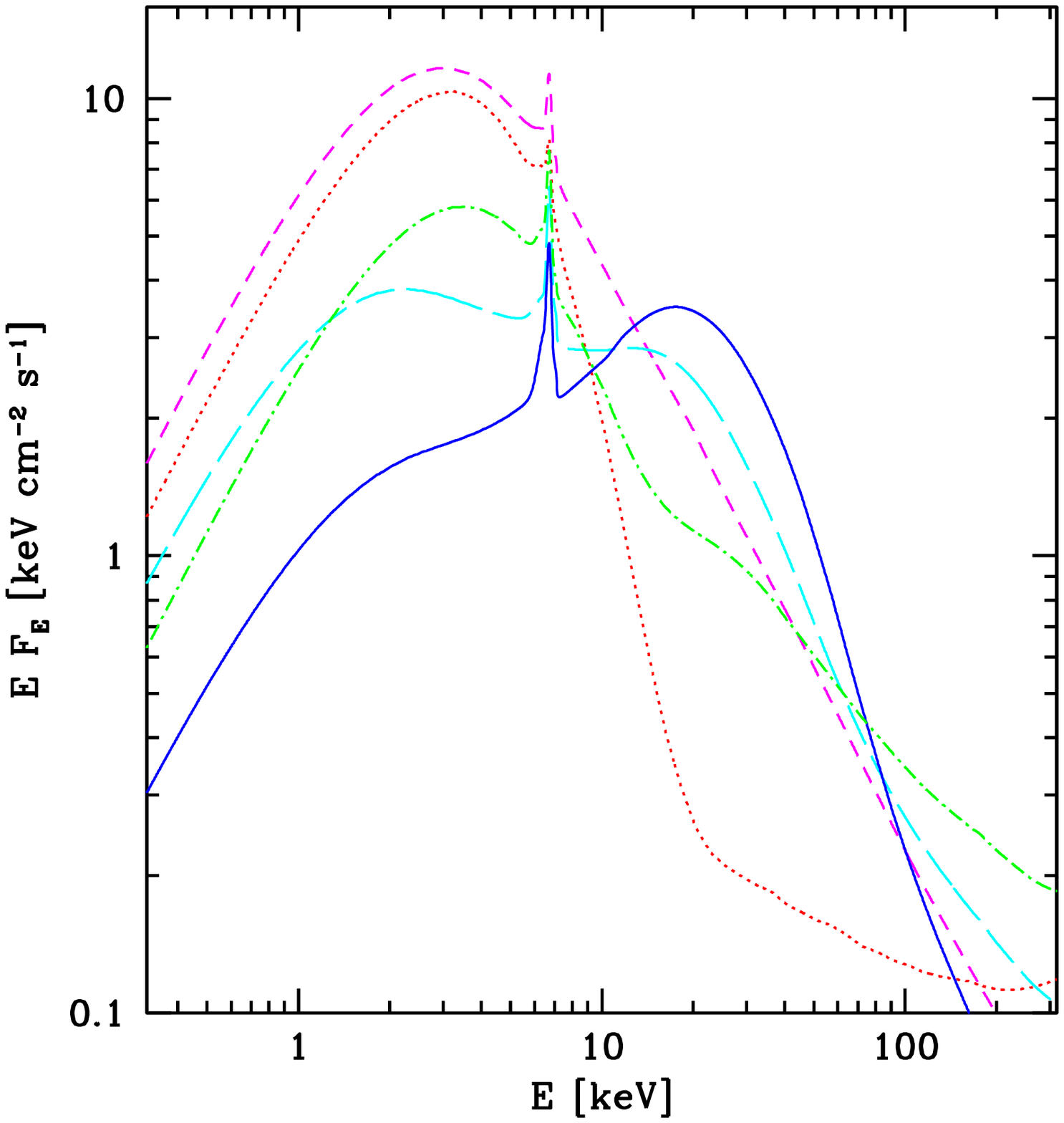} 
\includegraphics[width=0.45\textwidth,height=9cm]{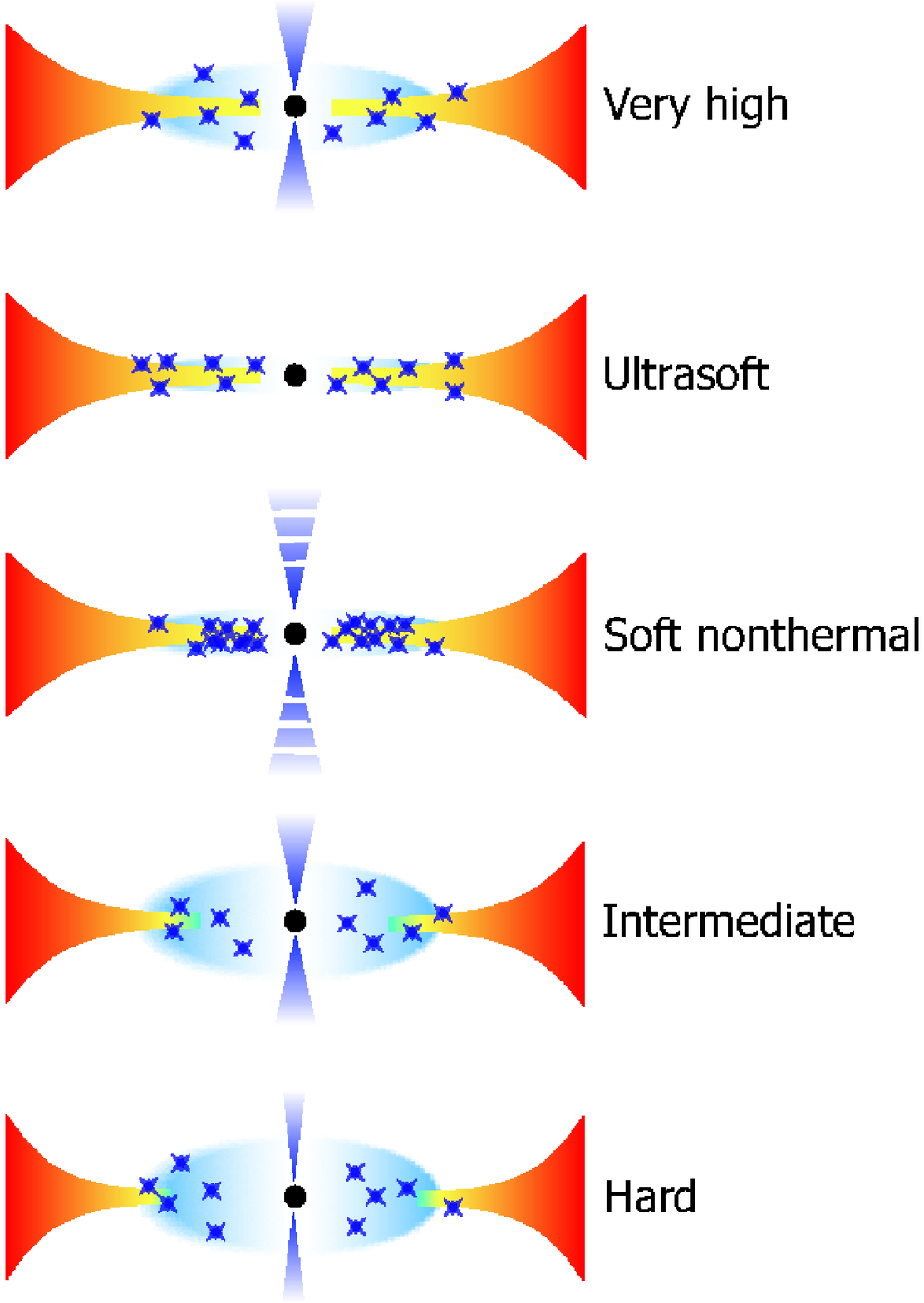} 
\caption{{\it a:} Spectral shapes of Cyg X-3 according to our models, corrected for absorption. The hard state (blue solid line), the intermediate state (cyan long dashes), the very high state (magenta short dashes), the soft non-thermal state (green dot-dashes) and the ultrasoft state (red dotted line). {\it b:} Proposed accretion geometries including radio behaviour. The non-thermal electron contribution is shown as blue stars.}
\label{unabs}
\end{figure*}

The unabsorbed spectral shapes according to our best fit models of the full range of the observed spectral variability are shown in Fig. \ref{unabs} together with the envisaged acccretion geometries.

\begin{table*}

\caption{The parameters of the best-fit models, as presented in Sections 4.1--4.5. The uncertainties correspond to 90 per cent confidence, i.e. $\Delta$\chisq$=+2.71$. The large error bars are a result of the presence of several themselves much more narrow minima.}
\begin{tabular}{lcccccccc}
\hline
 & 				  	&					{\rm hard state} &	{\rm intermediate state} &		{\rm very high state} &        	{\rm soft non-thermal} &	{\rm ultrasoft state}     \\
 \hline
%$N_{\rm H,gal}$& 	$10^{22}\, {\rm cm}^{-2}$ & 	$1.5{\rm f}$ &				$1.5{\rm f}$ &				$1.5{\rm f}$ & & \\
$N_{\rm He,0}$& 	$10^{22}\, {\rm cm}^{-2}$ &	$0.4^{+0.4}_{-0.4}$ &		$0.4^{+0.2}_{-0.4}$ &		$0.6^{+0.2}_{-0.6}$ &		$0.6^{+0.2}_{-0.6}$ &$0.7^{+0.1}_{-0.1}$ \\ 
$N_{\rm He,1}$& 	$10^{22}\, {\rm cm}^{-2}$ &	$1.7^{+1.7}_{-0.2}$ &		$0.8^{+1.3}_{-0.3}$ &		$1.9^{+2.7}_{-1.2}$ & 	$1.9^{+3.5}_{-1.7}$ & $0.2^{+0.1}_{-0.2}$ \\ 
$f_{1}$ &			 &						$0.6^{+0.2}_{-0.3}$ &		$0.5^{+0.3}_{-0.4}$ & 		$0.5^{+0.3}_{-0.3}$ &		$0.4^{+0.5}_{-0.2}$ &	$0.6^{+0.4}_{-0.6}$\\
%$N_{\rm He,scatt}$ &	$10^{22}\, {\rm cm}^{-2}$ & 	$3^{+0.00}_{-0.00}$ &	$3^{+0.00}_{-0.00}$ &		$3^{+0.00}_{-0.00}$ & $2.86^{+0.04}_{-0.02}$ &$3$ \\
${\rm A_{Fe}}$ & 	 &						$0.7^{+0.1}_{-0.2}$ &		$0.9^{+0.3}_{-0.1}$ &			$0.8^{+0.4}_{-0.1}$ & 	$0.4^{+0.2}_{-0.2}$ &	$0.9^{+0.6}_{-0.5}$\\
${\rm A_{>He}}$ &	&						$0.6^{+0.1}_{-0.1}$&		$0.7^{+0.2}_{-0.1}$&			$1.4^{+1.4}_{-0.7}$ & 	$1.0f$ &	$1.0f$\\
$E_{\rm edge}$ & 		${\rm keV}$&			$9.37^{+0.16}_{-0.14}$&		$9.28^{+0.06}_{-0.08}$ &  	$9.19^{+0.10}_{-0.09}$ & $9.14^{+0.09}_{-0.10}$&$9.14^{+0.07}_{-0.12}$\\
$\tau_{\rm edge}$ & 	 &						$0.17^{+0.02}_{-0.03}$ &		$0.25^{+0.02}_{-0.02}$ &		$0.27^{+0.02}_{-0.03}$ &	$0.27^{+0.03}_{-0.05}$ &	$0.34^{+0.03}_{-0.04}$\\
$kT_{\rm s}^{\rm a}$ & 		${\rm keV}$&		$0.6^{+0.3}_{-0.3}$ &		$0.6^{+0.9}_{-0.2}$&			$1.0^{+0.4}_{-0.3}$ &		$1.1^{+0.3}_{-0.2}$ &	$1.2^{+0.1}_{-0.1}$\\
$\lh/\ls$  &  		 &						$2.00^{+0.06}_{-0.15}$&		$0.90^{+0.16}_{-0.21}$&		$0.36^{+0.09}_{-0.05}$ &	$0.32^{+0.23}_{-0.19}$ &	$0.21^{+0.01}_{-0.04}$\\
$\lnth/\lh$ & 		 &						$0.38^{+0.11}_{-0.11}$&		$0.28^{+0.49}_{-0.06}$ &		$0.36^{+0.19}_{-0.19}$ &	$0.71^{+0.16}_{-0.09}$ &	$0.50^{+0.02}_{-0.02}$\\
$\Gamma_{\rm inj}^{\mathrm {b}}$ &	&					$3.9^{+0.1}_{-0.1}$&		$2.8^{+0.6}_{-0.3}$&			$3.5^{+0.4}_{-0.7}$ &		$2.5^{+0.4}_{-0.1}$ &	$2.1^{+0.1}_{-0.1}$\\	
$\tau_{\mathrm e}$ & 		 &				$4.76^{+0.12}_{-0.14}$&		$3.75^{+1.01}_{-0.39}$ &		$3.40^{+0.14}_{-2.69}$ &	$4.70^{+4.72}_{-1.51}$ &	$4.72^{+0.68}_{-0.26}$\\
$\tau_{\mathrm tot}^{\mathrm{c}}$ & 		 &		$4.77$&					$3.78$ &					$3.40$ &		$4.74$ &	$5.15$\\
$kT_{\mathrm e}^{\mathrm{c}}$ & ${\rm keV}$& 	8.3 &						7.6 &						4.6 &					2.2 		&	2.8\\
$R=\Omega/ 2\pi$ & 	 &					$2.5^{+1.5}_{-0.5}$ &		$2.2^{+2.5}_{-0.5}$ &			$2.3^{+1.0}_{-1.7}$ &		$2.0^{+5.9}_{-2.0}$ 	&$0.0^{+0.1}$\\
$E_{\mathrm Fe}^{\mathrm {d}}$ &			${\rm keV}$& 		$6.4^{+0.2}$ &		$6.4^{+0.2}$ &				$6.5^{+0.2}_{-0.1}$ & 	$6.5^{+0.1}_{-0.1}$	&$6.5^{+0.1}_{-0.1}$\\
$\sigma_{\mathrm Fe}^{\mathrm {e}}$ & 			 &			$0.3^{+0.1}_{-0.3}$ &		$0.4_{-0.2}$ &			$0.4_{-0.2}$ & 		$0.2_{-0.2}^{+0.2}$&	$0.4_{-0.1}$\\	
$EW_{\mathrm Fe}$	& ${\rm eV}$	&	$310$	&	$190$&	$120$&	 $200$&	$270$\\ 
$F_{\rm bol, abs}$ & 	${\rm erg\, cm}^{-2}\, {\rm s}^{-1}$ & $7.7\times 10^{-9}$ &	$7.9\times 10^{-9}$ &	$1.1\times 10^{-8}$ &$8.2\times 10^{-9}$  & $8.7\times 10^{-9}$\\
$F_{\rm bol}^{\rm g}$ & 	${\rm erg\, cm}^{-2}\, {\rm s}^{-1}$ & $1.6\times 10^{-8}$ &	$2.3\times 10^{-8}$ & 	$4.9\times 10^{-8}$ & $2.5\times 10^{-8}$&$3.7\times 10^{-8}$ \\
$L_{\rm bol}^{\rm h}$& ${\rm erg}\, {\rm s}^{-1}$& 	$1.6\times10^{38}$ &		$2.2\times10^{38}$ &		$4.7\times10^{38}$ &$2.5\times10^{38}$ & $3.6\times10^{38}$\\
$\chi^2/\nu$		 &					&	47/93 &					23/84 &					24/87 & 	45/74& 	60/89\\
\hline
\end{tabular}
\begin{list}{}{}
\item[$^{\mathrm{a}}$] For all states except the ultrasoft state, the best-fit blackbody temperature was frozen when determining the error ranges of the other parameters, see discussion in Section 3.3.  
\item[$^{\mathrm{b}}$] The acceleration power-law index $\Gamma$ for the non-thermal electrons was limited between 2.0 and 4.0.
\item[$^{\mathrm{c}}$] Calculated from the energy balance, i.e. not a free fit parameter.
\item[$^{\mathrm{d}}$] The energy of the additional broad iron line was limited between 6.4--7.5 keV.
\item[$^{\mathrm{e}}$] The physical width of the additional broad iron lime was limited to $\leq$ 0.4.
\item[$^{\mathrm{g}}$] The bolometric flux of the unabsorbed model spectrum normalized to the PCA data.
\item[$^{\mathrm{h}}$] The unabsorbed bolometric luminosity assuming a distance of 9 kpc (Dickey 1983, assuming 8 kpc for the distance to the Galactic Centre; Predehl et al. 2000).
\end{list}
\end{table*}

\section{Spectral evolution}

\subsection{The colour-colour diagram of Cyg X-3}
Having derived unabsorbed spectral shapes of the range of spectral states displayed by Cyg~X-3, we can now calculate its intrinsic X-ray `colours' and study the spectral evolution in a colour-colour diagram. To enable comparison with other classes of X-ray binaries, black holes as well as neutron stars, we use the same colours as Done \& Gierli\'nski (2003, hereafter DG03), where the soft colour is defined as the ratio of the intrinsic fluxes in the 4--6.4 keV and 3--4 keV bands, and the hard colour as the ratio of the intrinsic fluxes in the 9.7-16 keV and 6.4--9.7 keV bands. In Fig. \ref{cc}a the colour-colour diagram of Cyg~X-3 is plotted together with the data of the black holes and atoll sources from DG03, Gladstone, Done \& Gierli\'nski (2007) and Done, Gierli\'nski \& Kubota (2007). Note that the colours of Cyg X-3 are not corrected for reflection, i.e. they represent the observed although absorption corrected spectral shape and are therefore not dependent on the actual spectral decomposition of the unabsorbed spectra.

It is obvious from Fig. \ref{cc} that the track in the colour-colour diagram traced out by Cyg~X-3 does not resemble the characteristic Z-track of the atoll sources (the Z-like shape is actually more pronounced in the atolls than in the Z-sources, DG03; no Z-sources are included in Fig. \ref{cc}). It is rather more similar to that of the black holes, except that the the high values of the soft colour of Cyg X-3 in its soft states keep it from entering the lower left corner of the diagram, the area occupied {\it only} by black holes and inaccessible to neutron stars as defined by DG03. The high values of the soft colour in Cyg X-3 is due to the high blackbody temperature that results in a hard slope of the spectrum below 5 keV in the soft states. The soft states of Cyg X-3 occupy similar colours to those of GRS 1915+105, but it is obvious that Cyg X-3 spans a much wider range of hard colours than does GRS 1915+105, which does not show transitions into a hard state due to its constantly high luminosity, always exceeding $0.3\ledd$ (Done, Wardzi\'nski \& Gierli\'nski 2004, hereafter DWG04). Overall, the track of Cyg X-3 in the colour-colour diagram is rather similar to that of GRO J1655--40 (DG03, fig.\ 2), a black hole transient harbouring a 7$\msun$ black hole (Orosz \& Bailyn 1997) and, just like GRS 1915+105 and Cyg X-3, a strong radio source.

\subsection{The colour-luminosity diagram of Cyg X-3}
The mass of the compact object in Cyg X-3 is unknown. To study the spectral evolution as a function of $\ledd$, we thus have to assume a probable mass for the compact object. Alternatively, we may use the track traced out by Cyg X-3 in its colour-colour and colour-luminosity diagram to by comparison to other objects make an educated guess about the approximate mass and nature of its X-ray emitting object. 

The Eddington luminosity for the X-ray emitting compact object is 
\begin{equation}
L_{\rm E}\equiv 4\pi \mu_e GM m_p c/\sigma_{\rm T},
\end{equation}
where $\mu_e=2/(1+X)$ is the mean electron molecular weight, $X$ is the hydrogen mass fraction, and $\sigma_{\rm T}$ is the Thomson cross section. Assuming, despite the simplifications made in modelling of the absorption, that the accreted matter contains no or very little hydrogen, $\ledd$ is that of a helium-star, $\ledd=2.5\times 10^{38} M/\msun$ erg s$^{-1}$. 

\begin{figure*}
\includegraphics[width=0.45\textwidth]{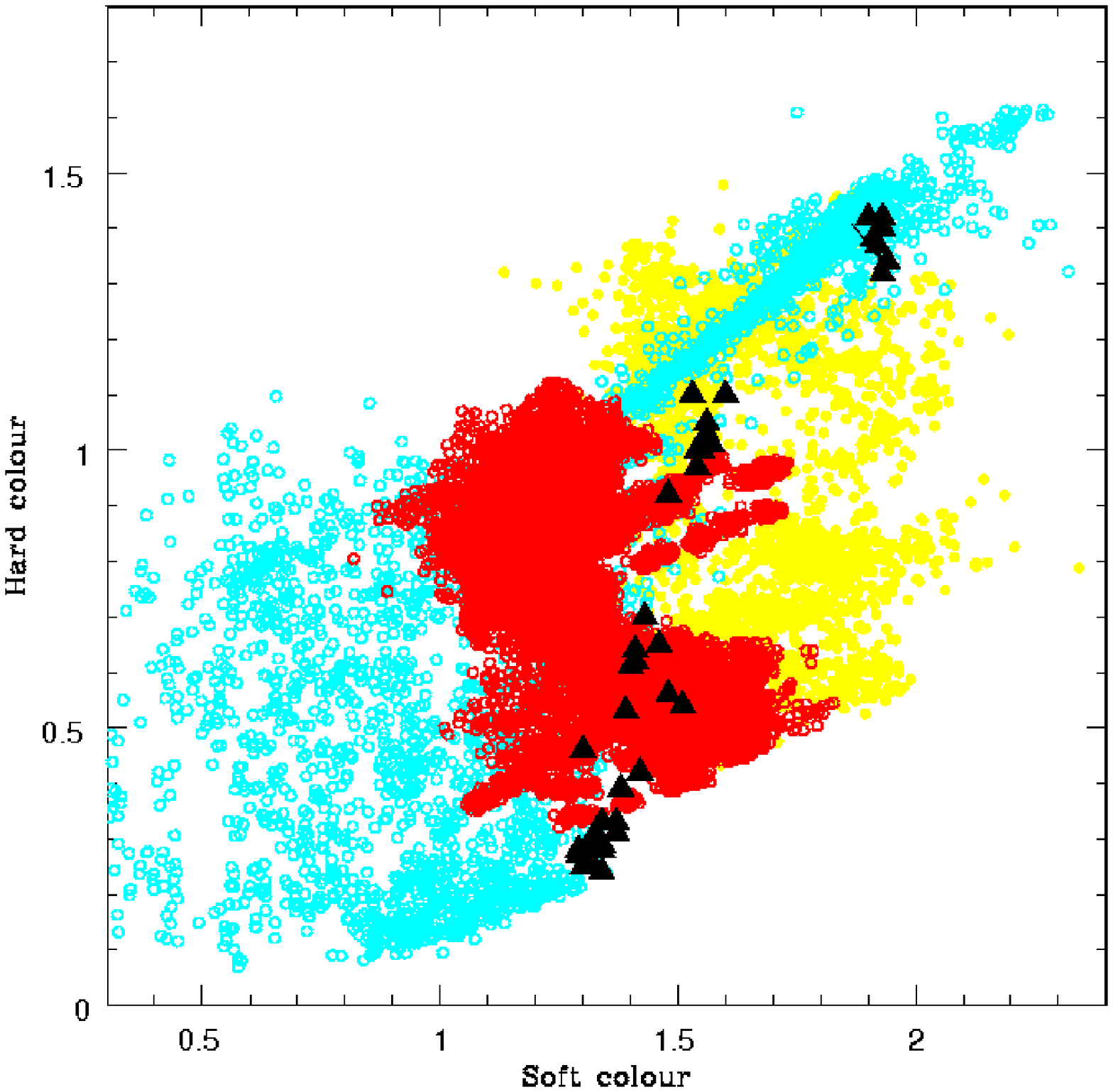}
\includegraphics[width=0.45\textwidth]{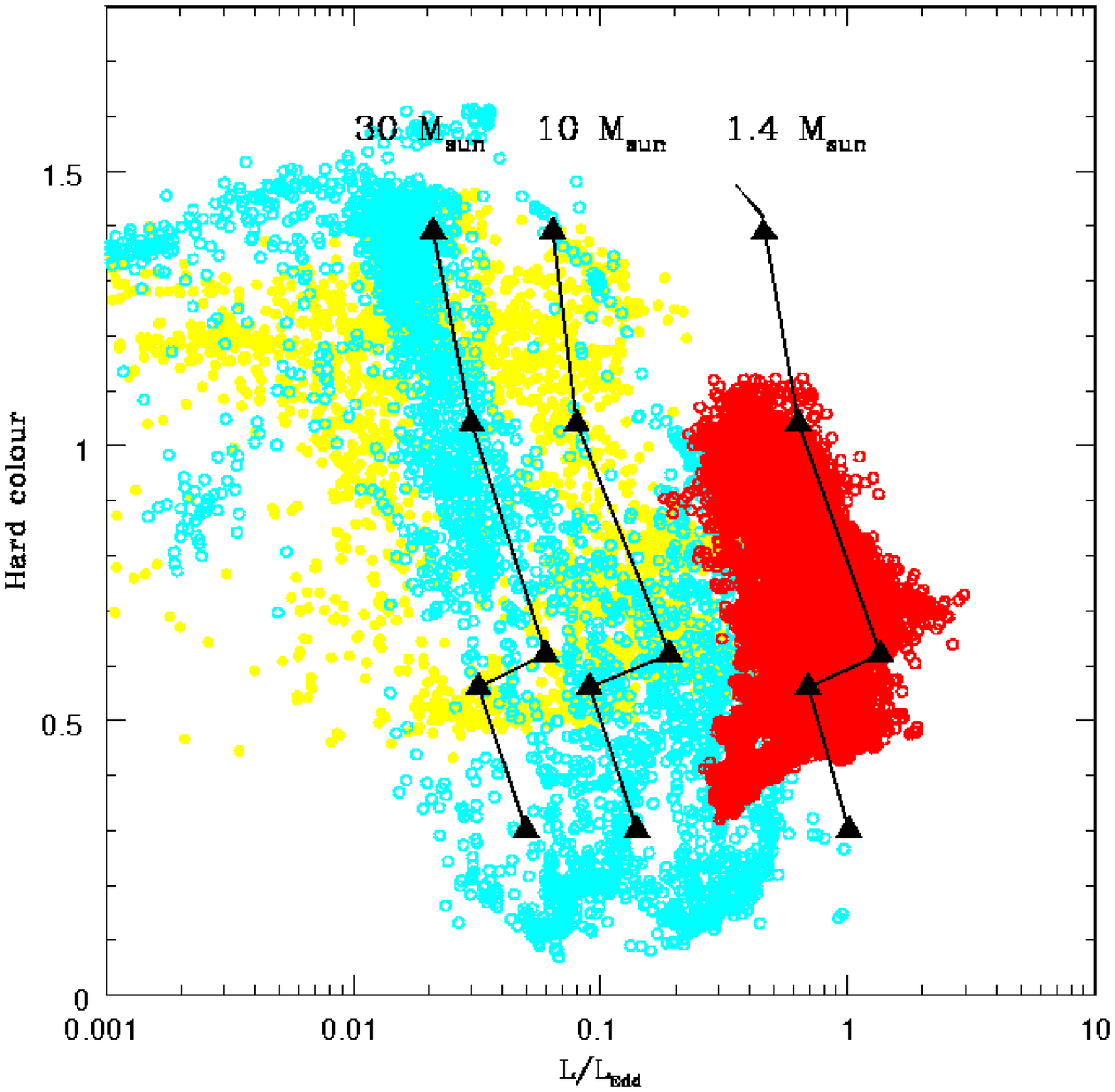}
\caption{{\it a:} Colour-colour diagram of Cyg X-3. All 42 individual observations are shown as black triangles overplotted on the data from the Galactic black holes (cyan circles) and atolls (yellow filled circles) from DG03, Gladstone et al. (2007) and Done et al. (2007), and GRS 1915+105 (red circles) from DWG04. {\it b:} Colour-luminosity diagram of Cyg X-3 assuming a mass of its compact object of 1.4, 10 and 30 $\msun$, from right to left (black triangles, only average points are shown for clarity). Data from the same Galactic black hole sources (cyan circles), atolls (yellow filled circles) and GRS 1915+105 (red circles) are overplotted.}
\label{cc}
\end{figure*}

In Fig. \ref{cc}b the hard colour is plotted against the luminosity from Table 2 as $L/\ledd$ for three different assumptions for the mass of the compact object in Cyg X-3. Only the results from the modelling of the average spectral shapes are plotted for clarity. Overplotted is the hard colour versus luminosity in Eddington units for the black holes and neutron stars (atolls) from DG03, Gladstone et al. (2007) and Done et al. (2007) (the same data as in Fig. \ref{cc}), for comparison. (Note that the lines connecting the points for the different mass-assumptions do not suggest the source evolution but are simply there to guide the eye).

The track to the right assumes $M_{\rm X}=1.4\msun$, representing a neutron star accretor. If the compact object is a neutron star, its path traced out in the CL-diagram differs significantly from that of the normal atoll sources. The luminosity is close to and sometimes exceeds $\ledd$. This behaviour is not observed in atolls but seems to be common in several Z-sources, neutron star X-ray binaries with constant high accretion rates (see e.g. DG03). As an extreme example, the Z-source Cir X-1 can reach highly superEddington luminosities and spans a range of $\la 10\ledd$ (DG03). These sources that are constantly accreting at high accretion rates {\it do not}, however, show state transitions. Their tracks in a colour-colour diagram span only a narrow range in hard colours, corresponding to a variable black-body temperature, and {\it never} displays hard colours exceeding 1.0, corresponding to the realm of hard states in the colour-colour diagram (see e.g. DG03, fig.\ 8). Thus, the fact that Cyg X-3 shows state transitions is incompatible with its compact object being a neutron star since state transitions to the hard state is not compatible with a constant high accretion rate.

The middle track is that for a compact object of $M=10\msun$, a typical mass for Galactic black holes. With this assumption, the luminosity span of Cyg X-3 corresponds to $\sim$0.06--0.2$\ledd$, comparable to some of the more luminous black hole systems, like eg. XTE J1550-564 and LMC X-3 (DG03, fig.\ 2). The transition to the hard state would however have to take place at $\sim0.1\ledd$. Such high transition luminosities are indeed observed in transient systems (e.g. $0.3\ledd$ for GX 339-4, Zdziarski et al. 2004), despite the fact that theoretical models of the advective solutions that are believed to describe the accretion flow in the hard state are stable only up to luminosities of a few per cent $\ledd$ (e.g. Esin et al. 1997). The fact that these high transition luminosities are observed only in transient systems, and only in the hard-to-soft transition, point to them being a result of the transient behaviour. When a transient enters an outburst, it starts from quiescence. If the local accretion rate and thereby the luminosity increases faster than there is time to build up an accretion disc (usually believed to take place on the viscous time scale), the source may stay in a hard state up to a higher luminosity than what would have been possible if an accretion disc was always present and the transition from a dominant hot advective to a cool disc solution was set solely by the properties of the respective solutions, irrespective of previous behaviour. The transient sources thus show hysteresis, i.e. the hard-to-soft transitions take place at higher luminosities than the soft-to-hard transitions. Cyg X-3, just like Cyg X-1, does not show any hysteresis. The transitions between the hard to soft and the soft to hard state always take place at the same luminosity. 
  
\begin{figure}
\includegraphics[width=0.5\textwidth]{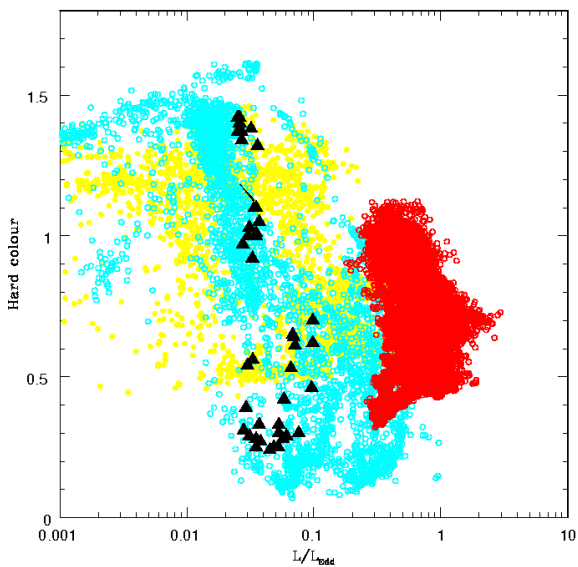}
\caption{The full colour-luminosity diagram for Cyg X-3 assuming a black hole mass of $30\msun$ overplotted on the same data from other black hole and atoll-sources as in Fig 8. About a factor of two in the spread in luminosity is due to orbital modulation.}
\label{cl30}
\end{figure}

For the transition to take place at a luminosity corresponding to $0.03\ledd$ which is the case in e.g. Cyg X-1 and LMC X-3 (DG03), both persistent systems like Cyg X-3, the mass of the compact object in Cyg X-3 would have to be as high as $30\msun$ (for $L$ to equal $0.03\ledd$ in its intermediate state). The track in the colour-luminosity diagram for Cyg X-3 assuming a compact object of $M=30\msun$ is shown to the left in Fig. \ref{cc}b. With this assumption, the luminosity span in Cyg X-3 corresponds to 0.02--0.07$\ledd$, comparable to that of Cyg X-1. The full colour-luminosity diagram for $M=30\msun$, including all 42 observations is shown in Fig. \ref{cl30}. The pattern is somewhat similar to that of LMC X-1 and GRO J1655 (DG03). Note that part of the spread in luminosity in the individual observations is due to orbital modulation. The difference in luminosity between observations at phase maximum and minimum differs by approximately a factor of 2.

We would like to point out that the values for the luminosity in Fig. \ref{cc}b are all lower limits. For the states where reflection exceeds 2.0, the implied intrinsic luminosity should be multiplied by a factor of minimum $R/2$ (assuming full sky coverage of the reflector as seen by the emitter) and maximum $R$ (for a slab-geometry) to account for the part of the emission obscured from view. (See further discussion in Hj08 and SZ08). If scattering in the stellar wind results in a net decrease of photons (as assumed in Hj08 and SZ08 but not here, see Section 3.2), this would also increase the intrinsic luminosity. The higher luminosity, the higher mass is needed to explain the data and our value for the required mass of the compact object of $30\msun$ is thus the most conservative one.  

\section{Discussion}
\subsection{State transitions and the presence of a hard state in Cyg X-3}
An important question regarding the spectral variability in Cyg X-3 has been whether it actually displays a state transition from a soft disc dominated state to a hard state with disc truncation in the same way as most other black hole binaries, or whether the observed spectral variability at energies $<10$ keV is caused by variable local absorption. 
%This was thoroughly investigated by Hjalmarsdotter et al. (2007), hereafter Hj07, who, based on a study of both the X-ray lightcurve and its correlation with the radio properties as well as spectral modelling of \integral\/ data, concluded that a state transition indeed takes place and represent an intrinsic re-configuration of the accretion flow. 
The absorbed spectral shapes resemble those of other black hole binaries (Szostek \& Zdziarski 2004), with the exception of the hard state that peaks at considerably lower energies than what is usually observed, as discussed in Section 4.1. The similarity of the spectral shape of the hard state at energies $>$20 keV, together with the presence of strong absorption, affecting all states up to higher energies than usual for normal Galactic absorption columns, may at first seem suggestive that the hard state is just a strongly absorbed version of one of the softer states (e.g. Hjalmarsdotter et al. 2004). Understanding the spectral variability in Cyg X-3 is thus very much dependent on understanding the nature of the hard state and assessing whether the transition to it marks a `real' state transition or not. The issue of the nature of the hard state of Cyg~X-3 was investigated thoroughly by Hj08. Results from a study of the X-ray lightcurve showed a bimodal distribution of the soft X-rays. This, together with the anticorrellation of soft and hard X-rays as well as between soft X-ray flux and hardness, is suggestive of a bimodal behaviour of the accretion flow, with intrinsic spectral pivoting rather than variable absorption being responsible for the observed state transitions. The similarity of the radio/X-ray correlation pattern in the hard state of Cyg~X-3 to that of other sources in the hard state further suggested that the accretion flow in the Cyg~X-3 hard state shares some common physics and/or geometry inherent to the canonical hard state. The fact that the orbital modulation was found to be stronger in the {\it soft} state than in the hard also speaks against the hard state being an artefact of increased wind absorption. These results led Hj08 to conclude that the hard state of Cyg~X-3 is likely to be a lower accretion rate state with a truncated or absent inner disc and that a state transition indeed takes place in Cyg X-3 and represents an intrinsic re-configuration of the accretion flow. 

\subsection{The location of the inner disc}
Together with the uncertainties about the exact properties of the absorber, the true blackbody temperature (and thereby the inner disc radius), especially in the hard state is hard to determine by spectral modelling. We would like to stress that this is not just a question of the high bandpass ($>$3 keV) of the \xte\/ data. The situation is almost equally problematic when using data extending to lower energies. When modelling the \sax\/ data in SZ08, the blackbody temperature in the hard state could only be determined to within $\Delta\chi^2/\nu$ of 0.14 for a range of 0.4--2.8 keV with the same phenomenological treatment of absorption as used here, and had to be frozen when modelling the wind parameters. The high blackbody temperature found as the best fit in SZ08 (1.7 keV) may well, as discussed in that paper, correspond not to the inner disc temperature but to some soft excess, often present in the spectra of black hole binaries (eg Cyg X-1, Gierli\'nski et al. 1997). In this paper, we find a model for the hard state consistent with a lower blackbody temperature (0.6 keV).% that can describe well the spectrum as observed by \xte but also the \integral\/ data in Hj08 and the \sax\/ data in SZ08, at least if abundances are allowed to differ from standard WR-abundances (which was not allowed in SZ08, and may not be corresponding to physical reality). 
 The exact black body temperature in the hard state remains, however, poorly known.

\subsection{Compton reflection and the geometry of the accretion flow}
A reflection dominated spectrum in the hard state of Cyg X-3 was first suggested by Hj08 as the best fit model to the \integral\/ data presented there. Also SZ08 found that a reflected dominated spectrum gave the best fit to \sax\/ data of Cyg X-3 in both its hard state and a state similar to the ultrasoft state presented here. In this paper, our best fitting models to four of the five observed states (all except the ultrasoft state) require strong reflection, exceeding $R=2.0$ (for $i=60\degr$), the maximum solid angle in an unobscured geometry. This implies that if the inclination is $>60\degr$ the direct emission from the inner accretion flow is to some extent obscured from view, even if the models presented here do not imply luminosity corrections for the obscured parts larger than the true luminosity being 1.25 times the observed one (as compared to a correction factor $>$5 in SZ08). Reflection domination is commonly observed in obscured AGN (e.g., Matt, Guainazzi \& Maiolino 2003). Recently, a Compton reflection dominated spectrum has also been suggested to explain a low non pulsating state of the neutron star binary GX~1+4 by Rea et al. (2005). A possible geometry is a thickened or perhaps warped disc. Such a geometry in the case of Cyg X-3 has to be symmetrical to the compact object, not to cause any significant spectral changes related to the orbital phase, which are not observed.

If the system inclination is lower, the observed strength of the Compton reflection component corresponds to a smaller value of $R$, and no corrections to the luminosities given in Table 2 are required. The upper limit on the inclination from the lack of eclipses for the two extreme binary models considered in SZ08 is $77\degr$ and $64\degr$ for a companion star mass of 5 and $50\msun$, respectively. The higher value for the low mass of the companion star corresponds to a neutron star accretor, which we rule out. Since the star has an extreme stellar wind that is optically thick near the surface, the actual inclination has to be substantially smaller and is probably $\leq60\degr$.  

Even if the presence of a strong reflection component can explain the shape of the hard state, the unusually low electron temperature in the hard state, however, remains to be explained. The universal cut-off at $\sim100$--200 keV observed in both Galactic black holes (Zdziarski \& Gierli\'nski 2004) and in Seyfert I type AGN (Zdziarski et al. 1996a), strongly suggests that the hard state is usually characterised by electron temperatures of the order of 50--100 keV. A tentative explanation is that the low electron temperature as well as the unusually high fraction of non-thermal electrons in the hard state is somehow a result of the turbulent environment with the accretion flow enshrouded in the strong Wolf-Rayet wind. If the accreted energy is transferred to the electrons in the form of acceleration rather than thermal heating, this could be a sign of constant shocks and collisions between the wind and the accretion flow, as discussed in Hj08. A highly efficient acceleration process may also be responsible for the unusually strong radio emission in Cyg X-3, even if this can also be explained simply by the high density surrounding medium increasing the radiative efficiency of the jet. 

An alternative scenario that would result in a similar shape of the hard state spectrum is down-scattering of initially much higher temperature photons in a dense wind. The optical depth of such a wind has to be around $\tau\sim3-5$ in order to down-scatter photons with initial energies of 50--100 keV down to the observed cut-off at $\sim20$ keV. This high optical depth can probably not be achieved in the surrounding Wolf-Rayet wind, which was found by SZ08 to have an optical depth of $\le1$. A dense disc wind instead of a thickened disc could however cause down-scattering instead of single reflection in a similar geometry. While such a scenario could explain the low electron temperature, it would imply a much stronger initial high-energy luminosity than the one calculated from our models which would point to an even higher mass of the compact object than the one suggested here.

\subsection{The mass of the compact object}
Our results favour a massive black hole accretor in Cyg X-3 and the presence of a neutron star could be ruled out due to the high luminosity at the observed state transitions. A value of $M=30\msun$ corresponds to the state transition taking place at $L=0.03\ledd$,  in agreement with values observed in other sources and consistent with our present theoretical models for advective flows. We stress that our given values for the luminosities are the most conservative ones with respect to our models. Firstly, we do not correct for any net loss of photons by scattering in the wind due to orbital modulation (Section 3.2). Secondly, we do not correct the luminosity for the part of the emission obscured from our line of sight by the reflector (Section 6.2). The combined effect of correcting for both these effects may increase the given luminosities by up to a factor of 1.7.   

A massive back hole is consistent with the results of Schmutz, Geballe \& Schild (1996) who derive a mass for the compact object of 7--40$\msun$ for a range of Wolf-Rayet masses from 5 to 20$\msun$ and inclinations 30--90$\degr$, for  the mass function of $2.3\msun$. Limiting the inclination to $\leq60\degr$ and allowing a Wolf-Rayet mass between 5--50$\msun$, this mass function gives a mass of the compact object of 9--60$\msun$. A black hole mass of 30$\msun$ would imply a Wolf Rayet star mass of $\leq57\msun$. A lower inclination allows for a lower Wolf-Rayet mass.   

Interestingly, recent results from mass measurements of the only two other Wolf-Rayet X-ray binaries discovered so far, IC 10 X-1 (Prestwich et al. 2007, Silverman \& Filippenko 2008) and NGC 300 X-1 (Carpano et al. 2007a, b), suggest that the compact object in both cases are black holes and in at least one of them, IC 10 X-1, the most probable mass is within 23--34$\msun$.

\section{Conclusions}
We have modelled Cyg X-3 in all spectral states as observed by \xte\/, and for the first time derived unabsorbed spectral shapes and luminosities. Our modelling suggest that in 4 of the 5 spectral states, Compton reflection exceeds $R=2$, corresponding to the maximum solid angle in an unobscured geometry, indicating that the emission from the innermost accretion flow is partly obscured (for $i\ga60\degr$). We suggest that the peculiar shape of the spectrum in the hardest state of Cyg X-3 is caused by this strong Compton reflection on top of an intrinsic moderately hard state with a truncated disc, where the plasma electron temperature is much lower and the non-thermal electron fraction much higher than what is usually observed in other systems. The low electron temperature could be due to down-scattering in a dense disc wind. The soft states are, corrected for absorption, very similar to those displayed by other sources like GRO J1655-40 and GRS 1915+105.   

We discuss the spectral evolution of Cyg~X-3 and compare with other sources, using colour-colour and colour-luminosity diagrams. In our interpretation of the intrinsic absorption corrected spectra, Cyg X-3 spans the whole range of spectral states as seen in other black hole sources. This is also consistent with its radio behaviour (Szostek et al. 2008). We find that the track traced out by Cyg X-3 in the colour-colour diagram is very different from that of the neutron star atoll sources but similar to that of black hole binaries, even if the Cyg X-3 colours do not enter the 'black hole only' regime as defined by DG03. Based on our most conservative estimates of the intrinsic luminosities, we study the spectral evolution as a function of $\ledd$, assuming three different masses for the compact object. We find that a neutron star accretor is not a likely candidate since not only the brightest states but also the transition to the hard state would then take place at luminosities close to Eddington, which is neither observed in other sources nor supported by our present theoretical understanding. For the state transition to take place at the expected $L\simeq0.03\ledd$, the compact object must be a black hole of $M\simeq30\msun$, similar to the Wolf-Rayet black hole binaries recently discovered in other galaxies.

\section*{Acknowledgements}
This research has made use of data obtained through the High Energy Astrophysics Science Archive Research Center (HEASARC) Online Service, provided by NASA/Goddard Space Flight Center. We thank Marek Gierli\'nski for providing the data in Fig. \ref{cc}. DCH gratefully acknowledges a Fellowship from the Finnish Academy. LH acknowledges support from the Finnish Academy project nr 1118854, and from the Finnish Academy of Science and Letters, Vilho, Yrj\"o and Kalle V\"ais\"al\"a foundation. AAZ, AS and LH have been supported in part by the Polish MNiSW grant NN203065933 (2007--2010) and the Polish Astroparticle Network 621/E-78/SN-0068/2007. In its early stages this work was also partly funded by NORDITA through the Nordic Project in High Energy Astrophysics.

\label{lastpage}
\end{document}